\documentclass[altaffilletter,aps,nofootinbib,twocolumn,prd,eqsecnum,preprintnumbers,superscriptaddress,10pt,floatfix]{revtex4-1}
\pdfoutput=1
\usepackage{graphicx}
\usepackage{amsmath}
\usepackage{subfigure}
\usepackage{amssymb}
\usepackage{amsfonts}
\usepackage{mathtools}
\usepackage{amssymb}
\usepackage{enumerate}
\usepackage{xcolor}
\usepackage{bm}
\usepackage{mathrsfs}
\usepackage{epstopdf}
\usepackage{url}
\usepackage{footnote}
\usepackage{textcomp}
\usepackage{dsfont}
\usepackage{ulem}
\usepackage{hyperref}
\usepackage{enumerate}   
\usepackage{appendix}
\usepackage{textcomp}
\usepackage{tipa}

\makeatletter
\newcommand*{\rom}[1]{\expandafter\@slowromancap\romannumeral #1@}
\makeatother

\begin{document}

\title{Imaging a semi-classical horizonless compact object with strong redshift}

\author{Che-Yu Chen}
\email{b97202056@gmail.com}
\affiliation{RIKEN iTHEMS, Wako, Saitama 351-0198, Japan}

\author{Yuki Yokokura}
\email{yuki.yokokura@riken.jp}
\affiliation{RIKEN iTHEMS, Wako, Saitama 351-0198, Japan}

\begin{abstract}
The recent advancements in black hole imaging have opened a new era of probing horizon-scale physics with electromagnetic radiation. However, a feature of the observed images, a bright ring encircling a relatively dark region, has not sufficiently proved the existence of event horizons. It thus requires extreme care when studying the possibility of using such image features to examine quantum effects that may change the classical picture of black holes slightly or drastically. In this work, we investigate the image of a horizonless compact object, whose interior metric satisfies the 4D semi-classical Einstein equation non-perturbatively for the Planck constant, and whose entropy agrees with the Bekenstein-Hawking formula. Although the absence of an event horizon allows light rays to pass through the dense interior, the extremely strong redshift significantly darkens the image, making it almost identical to the classical black-hole image. In particular, if there is light emission a bit inside the surface of the object, the intensity around the inner shadow is slightly enhanced, which could be a future observable prediction to characterize the object. We also find through a phenomenological parameter that the image is further darkened due to interactions inside. Thus, the image is consistent with current observations, and the object could be a candidate for black holes in quantum theory.

\end{abstract}

\maketitle

\section{Introduction}\label{sec.intro}

One prediction in General Relativity is the gravitational lensing effects, i.e., light rays are bent when passing around a massive object, which are important in the contexts of gravitational physics \cite{Cunha:2018acu}, astrophysics \cite{Bartelmann:2010fz,Wambsganss:1998gg}, and cosmology \cite{Hoekstra:2008db}. In particular, the lensing effects around a black hole are so strong that the images of a black hole illuminated by its surrounding light emission are very different from those of usual massive objects. More explicitly, the photon trajectories can be extremely curved such that the light rays can orbit the black hole multiple times, or even infinite times, when propagating near it. The infinitely lensed photons correspond to a set of unstable spherical photon orbits around the black hole. The observed black hole image is thus characterized by a bright ring that consists of the intensity of lensed photons, on top of the foreground direct emission \cite{Gralla:2019xty}. On the other hand, the photons entering the event horizon cannot be accessed from outside. Therefore, the bright ring on the image plane would encircle a dark region that corresponds to photon trajectories terminating at the horizon. A bright ring encircling a region with central brightness depression, i.e., the shadow, is an important feature of the images of a black hole in (classical) GR \cite{Luminet:1979nyg,Falcke:1999pj,Bozza:2010xqn,Cunha:2018acu}. The recent progress in imaging supermassive black holes \cite{EventHorizonTelescope:2019dse,EventHorizonTelescope:2022wkp} achieved by the Event Horizon Telescope collaborations has confirmed such a feature.

The lensing scenario discussed above is for classical black holes. It is well known that GR still has its own issues, e.g. existence of singularities, incompatibility with quantum theories, etc, whose resolution may require quantum gravitational corrections. In particular, one unresolved puzzle of the quantum nature of black holes is the information paradox \cite{Hawking:1976ra}, which indicates that the identity of black holes consistent with quantum theory is still unknown. Essentially, the problem is related to the presence of horizons \cite{Mathur:2009hf,Raju:2020smc}, and therefore, the possibility that large quantum corrections may appear at the horizon scale and even avoid the formation of a horizon \cite{Mazur:2001fv,Mathur:2005zp,Barcelo:2007yk,Almheiri:2012rt,Kawai:2013mda,Barcelo:2015noa}, 
should be seriously considered. It thus becomes crucial to investigate whether one can probe these horizon-scale quantum effects via the shadow images of these quantum-corrected black holes \cite{Held:2019xde,Liu:2020ola,Brahma:2020eos,Afrin:2022ztr,Eichhorn:2022oma,Zhang:2023okw}.

Theoretically, a horizonless compact object can also generate a ring-like structure in its image as long as it is sufficiently compact such that unstable photon orbits exist. The most distinctive feature when there is no event horizon is that some light rays may be reflected at the surface of the compact object, or may propagate through its interior (see Ref.~\cite{Cardoso:2019rvt} for a review about testing horizonless compact objects via observations). If these photons pick up additional intensities during their propagation and reach the observer, they may contribute extra rings inside the major bright ring on the image, hence enhancing the central brightness. Such features have been identified in the models of wormholes \cite{Ohgami:2015nra,Shaikh:2018oul,Paul:2019trt,Vincent:2020dij,Wielgus:2020uqz,Wang:2020emr,Tsukamoto:2021caq,Peng:2021osd,Olmo:2021piq,Eichhorn:2022fcl,Guerrero:2022qkh,Delijski:2022jjj,Olmo:2023lil}, boson stars \cite{Cunha:2017wao,Olivares:2018abq,Rosa:2022tfv}, gravastars \cite{Sakai:2014pga}, fluid stars \cite{Rosa:2023hfm}, fuzzballs \cite{Mayerson:2023wck}, and naked singularities \cite{Shaikh:2018lcc,Gyulchev:2019tvk,Shaikh:2019hbm,Gyulchev:2020cvo,Stashko:2021lad,Stashko:2021het,Tsukamoto:2021fsz,Huang:2024bbs}. Improving future dynamic range, which, roughly speaking, quantifies the level of difference between the brightest and dimmest image features, may help to distinguish such horizonless compact objects from a black hole covered by a horizon \cite{Carballo-Rubio:2022aed}.

However, when the photons are allowed to enter the object, some subtle effects inside have to be carefully considered. These include the strong redshifts that can retard the photon propagation as seen by a distant observer, as well as possible interactions between light rays and the internal quantum structure of the object. Intuitively, the combination of these two factors will effectively \textit{block} the photons propagating inside the object, hence suppressing the intensity of the inner rings, and resembling more the classical black hole images. Modeling the internal interactions and their associated darkening effects is challenging and is often model-dependent. Recently, an effective approach to darkening images was proposed in the construction of the images of fuzzballs \cite{Bacchini:2021fig} and compact topological solitons \cite{Heidmann:2022ehn}. It was shown that the shadow images of these horizonless compact objects can be almost indistinguishable from classical black hole images. Gravitational redshifts, on the other hand, can darken images in a more model-independent manner because gravity is determined by a given energy-momentum distribution, independent of the details of the microscopic components. Therefore, the presence of extremely strong redshift should darken the images of a horizonless compact object in a way that does not depend much on the details of its internal interactions. In this paper, we will examine how generic such a darkening process is and how much the darkened images mimic the images of a classical black hole.

One way to find a candidate of black holes in quantum theory is to identify the most compact object formed in the time evolution of a collapsing matter according to the 4D semi-classical Einstein equations \cite{Kawai:2013mda,Kawai:2014afa,Kawai:2015uya,Ho:2016acf,Kawai:2017txu,Kawai:2020rmt,Kawai:2021qdk}
\begin{equation}
G_{\mu\nu}=8\pi G \langle \psi|T_{\mu\nu}|\psi\rangle\,.
\label{Einstein}
\end{equation}
This is the self-consistent equation in a mean-field approximation of quantum gravity, where gravity is described by a classical metric $g_{\mu\nu}$, and matter is represented by quantum operators \cite{Birrell:1982ix,Kiefer}. In this framework, a macroscopic object is described as a collection of many excited quanta in an excited state $|\psi\rangle$. In Ref.~\cite{Kawai:2013mda,Kawai:2014afa,Kawai:2015uya,Ho:2016acf,Kawai:2017txu,Kawai:2020rmt,Kawai:2021qdk}, by solving Eq.~\eqref{Einstein} self-consistently, a 4D spherically symmetric spacetime region was obtained as the configuration formed in the collapse of a spherical matter, where the backreaction of particles created during the collapse was considered. It represents a compact object with an outer surface $R_\textrm{out}$ located just outside the Schwarzschild radius $a_0(\gg l_p\equiv\sqrt{\hbar G})$ instead of a horizon, and it evaporates due to Hawking-like radiation in the timescale $\Delta t \sim a_0^3/l_p^2$. For a generic collapse, a dense structure with an exponentially large redshift and near-Planckian curvatures is formed around the surface, while the structure in deeper regions depends on the details of the initial distribution of the collapsing matter \cite{Kawai:2015uya}. For an adiabatic formation in a heat bath at Hawking temperature, the dense structure continues inward (except for a small center part) to form a radially uniform dense configuration, which can be considered the most thermodynamically typical \cite{Kawai:2014afa,Kawai:2021qdk}. In this paper, we focus on the uniform dense configuration and dub it the \textit{quantum horizonless compact object (QHCO)}. In addition, one can evaluate the thermodynamic entropy of the interior by a thermodynamic method  \cite{Kawai:2015uya,Yokokura:2022phq} and a statistical-mechanical one \cite{Kawai:2020rmt}, to reproduce the Bekenstein-Hawking formula \cite{Bekenstein:1973ur,Hawking:1975vcx}.\footnote{Here, the strong self-gravity plays an essential role in changing the entropy from the volume law to the area law \cite{Yokokura:2022phq}.} It was later found \cite{Yokokura:2023wxp} that the QHCO is the entropy-maximized configuration that saturates the entropy bounds proposed by Bekenstein \cite{Bekenstein:1980jp} and Bousso \cite{Bousso:1999xy}. Thus, QHCO has some properties that a black hole should have in quantum theory and could be a candidate for a quantum black hole.

We would like to emphasize again that the QHCO model satisfies self-consistently the semi-classical Einstein equations \eqref{Einstein} with many matter fields.\footnote{\label{foot:solution}Indeed, one can use a technique based on the properties of the interior metric, 
evaluate the renormalized energy-momentum tensor $\langle\psi|T_{\mu\nu}|\psi\rangle$ for, say, many massless scalar fields, and equate it to $G_{\mu\nu}$, to solve Eq.~\eqref{Einstein}
self-consistently \cite{Kawai:2020rmt}.
As a result, the relation \eqref{sigma} is obtained.} The high curvature inside induces quantum fluctuations of quanta, generating a large tangential pressure $\langle T^\theta{}_\theta \rangle$
\cite{Kawai:2020rmt} that supports the configuration against the strong self-gravity.\footnote{This is precisely why QHCOs can evade the Buchdahl limit \cite{Buchdahl:1959zz} and reach extreme compactness, similar to the case of very anisotropic stars \cite{Raposo:2018rjn}.} Although the model obeys the semi-classical description, it non-trivially represents a non-perturbative solution in $\hbar$ in the sense that the curvatures and the metric itself cannot exist in the limit $\hbar\rightarrow0$. Indeed, it belongs to a branch in the solution space of Eq.~\eqref{Einstein} different from those that include classical black hole metrics \cite{Kawai:2021qdk,Ho:2023eem}.

Of course, the semi-classical description naturally breaks down when the energy scales become Planckian. In the QHCO model, as a result of the balance between the pressure and gravity, the excited quanta spread over the whole interior (except for the central region), rather than concentrating near the center. Hence, the curvatures remain finite $\sim 1/n l_p^2$, which are non-perturbative in $\hbar$ but still in the semi-classical regime if the number of the degrees of freedom $n$ in the theory is large but finite: $n=O(1)\gg1$.\footnote{Here, $O(1)$ means $O(a_0^0)$ for $a_0\gg l_p$.}
On the other hand, the small central region has only a small energy $\sim m_p\equiv\sqrt{\hbar/G}$ and is not excited enough to be described semi-classically. Therefore, its complete description would require a self-consistent quantum theory of gravity. However, since it has only small energy, it should not involve a drastic nature of quantum gravity, and it is natural to assume that the central region of size $\sim\sqrt{n}l_p$ can be approximately described by a flat spacetime \cite{Kawai:2020rmt,Ho:2023eem,Yokokura:2022phq,Yokokura:2023wxp}. As a consequence, no singularity exists in the whole space.

The main theme of this paper is to investigate the images of QHCOs neglecting the time evolution of the evaporation. As we will show later, light rays are allowed to propagate through the interior of QHCOs due to the absence of the event horizon. If these light rays can reach the observer, they can generate additional bright rings inside the major ring on the images. However, in reality, these light rays will take an extremely long time during their passage in the dense configuration due to the exponentially strong redshift. The delays of these light rays then effectively darken the observed images, such that the darkened images become almost indistinguishable from those of classical black holes. We expect this feature to be insensitive to the details of possible internal interactions. In fact, the QHCO metric incorporates the internal interaction effects through a phenomenological parameter $\eta$ (see sec.~\ref{sec.meco}), and we show that the image becomes closer to that of classical black holes for stronger interactions. These render the QHCO a perfect model for black hole mimickers. In addition, we find that a tiny feature of the QHCO image, which is more pronounced for larger $n$, will appear if light sources exist just inside the QHCO surface. This prediction is independent of the details of the emission profile, as long as the inner part of the surface is lit up. Therefore, such a feature may be a test-bed for near-horizon quantum effects with future improvements in observational techniques, e.g., substantially high dynamic range.

This paper is organized as follows. In sec.~\ref{sec.meco}, we review the QHCO spacetime more explicitly, including the metric and some properties of QHCOs. In sec.~\ref{sec.photoneq}, we investigate the photon geodesics of QHCO spacetimes. Considering a QHCO surrounded by an optically and geometrically thin accretion disk, we investigate the ``ideal" QHCO images in sec.~\ref{sec:image} assuming that the light rays propagating through QHCOs can reach the observer, no matter how long they may take inside QHCOs as seen by the observer. Then, in sec.~\ref{sec:image2}, we generate a set of more realistic QHCO images by including the darkening effects caused by the strong redshifts inside QHCOs. We discuss how the images may differ from, or mimic, the images of a classical Schwarzschild black hole. Finally, in sec.~\ref{sec:conclusion}, we conclude the paper by summarizing the results obtained and discussing their implications to fundamental aspects of quantum gravity.

\section{Quantum horizonless compact object}\label{sec.meco}
\begin{figure}[h]
  \centering
 \includegraphics[scale=0.95]{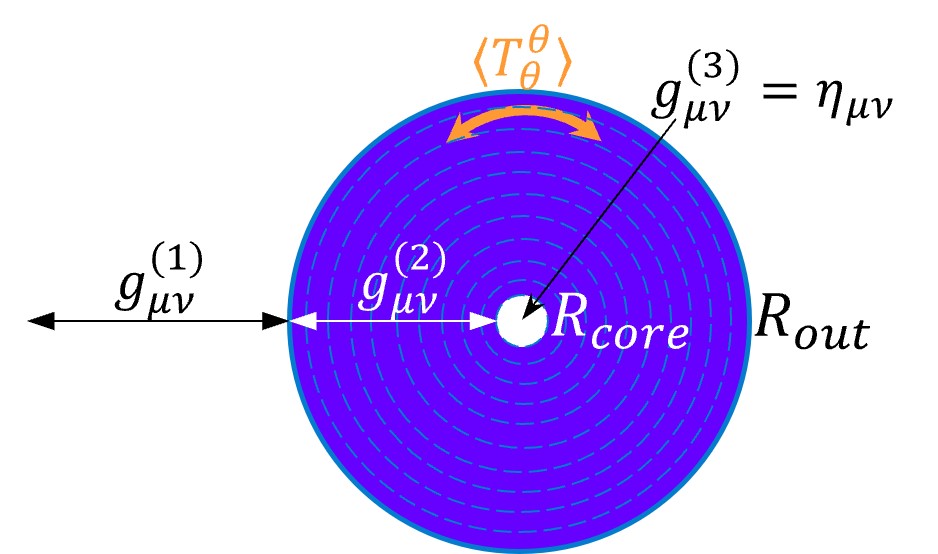}
\\
\caption{The configuration of the QHCO model. It consists of three regions and has inner and outer surfaces. The excited quanta inside are supported by the large quantum pressure $\langle T^\theta{}_\theta \rangle$ against the self-gravity so that they distribute uniformly in the radial direction.}
\label{fig:model} 
\end{figure}

As mentioned in the Introduction, the QHCO has an outer surface at $r=R_\textrm{out}$. Inside the surface, a collection of highly excited quanta forms a dense configuration with large curvatures and exponentially strong redshifts. On the other hand, the small central core is approximated by a flat space, whose size $R_\textrm{core}$ is comparable to the size of a quantum bit $\sim\sqrt{n}l_p$.  
Therefore, these surfaces separate three different spacetime regions as shown in Fig.~\ref{fig:model}.

More explicitly, the QHCO spacetime can be described by the following spherically symmetric and static metric \cite{Kawai:2013mda,Kawai:2014afa,Kawai:2015uya,Kawai:2017txu,Kawai:2020rmt,Kawai:2021qdk,Yokokura:2022phq,Yokokura:2023wxp}:
\begin{equation}
ds^2=-A_{i}(r_i)dt_{i}^2+B_{i}(r_i)dr_i^2+r_i^2d\Omega_2^2\,,\label{metric}
\end{equation}
where the subscripts $i=1$, $2$, and $3$ label the quantities in the exterior region, the dense configuration, and the central core, respectively.\footnote{Note that, for convenience, the radial coordinates in each region are written as $r_i$, but they all belong to the same coordinate $r$.}

The spacetime outside the surface $R_\textrm{out}$ is approximately described by the Schwarzschild metric:\footnote{In this metric, for simplicity, we neglect a small backreaction from vacuum polarization \cite{Birrell:1982ix}. Its possibly large effect near the Schwarzschild radius is considered in the interior metric \eqref{dense_metric}.} 
\begin{equation}
A_1\left(r_1\right)=\frac{1}{B_1\left(r_1\right)}=1-\frac{a_0}{r_1}\,.
\end{equation}

The dense configuration is expressed as 
\begin{equation}
A_2\left(r_2\right)=\frac{2\sigma}{r_2^2}\,\textrm{exp}\left({\frac{r_2^2}{2\sigma\eta}+C}\right)\,,\qquad B_2\left(r_2\right)=\frac{r_2^2}{2\sigma}\,,
\label{dense_metric}
\end{equation}
where the constant $C$ will be fixed below. There are two parameters $(\sigma,\eta)$. 
Physically, $\sigma=O(n l_p^2)$ represents the intensity of Hawking-like radiation \cite{Kawai:2013mda,Kawai:2014afa}. On the other hand, the parameter $\eta$ is a dimensionless constant satisfying 
\begin{equation}
    1\le\eta<2\,,
    \label{eta}
\end{equation}
which comes from the positivity of pressures and the causality of the interior matter \cite{Kawai:2020rmt}. Microscopically, $\eta$ depends on internal interactions in the sense that $\eta=1$ corresponds to radial lightlike propagation without scattering while $\eta\neq 1$ has effects of interactions \cite{Kawai:2014afa}. This is reflected in the fact that a larger $\eta$ leads to a smaller radial pressure 
$\langle T^r{}_r(r)\rangle$, as one can see from the energy-momentum tensor: 
\begin{align}
\langle -T^t{}_t(r)\rangle=\frac{1}{8\pi G r^2}&\,,~
\langle T^r{}_r(r)\rangle=\frac{2-\eta}{\eta}\langle - T^t{}_t(r)\rangle\,,\nonumber\\
\langle T^\theta{}_\theta(r)\rangle&=\frac{1}{16 \pi G \sigma \eta^2}\,,
\label{EMT}
\end{align}
which are the leading-order expressions for $r\gg l_p$ when applying the metric \eqref{dense_metric} to Eq.~\eqref{Einstein}. 
Note that the tangential pressure is near Planckian, i.e., $\langle T^\theta{}_\theta(r)\rangle\sim 1/G n l_p^2$ (from Eq.~\eqref{sigma}), and one cannot take the classical limit $\hbar\to0$. As a result, the dominant energy condition is violated, and the interior is locally anisotropic.

The parameters $(\sigma,\eta)$ can be determined by solving Eq.~\eqref{Einstein} self-consistently for a given theory satisfying the condition \eqref{eta} \cite{Kawai:2020rmt} (see also footnote \ref{foot:solution}). In particular, we have a relation 
\begin{equation}
    \sigma=\frac{f n l_p^2}{\eta^2}\,,
    \label{sigma}
\end{equation}
where $f$ is a numerical coefficient. As a result, the intensity $\sigma$ of the Hawking-like radiation is proportional to the number $n$ of the degrees of freedom in the theory.

Now, we fix the constant $C$ by using Israel's junction condition \cite{Poisson_book}, that is, the continuity of the metric functions $A_1(r_1)$ and $A_2(r_2)$ at the surface $R_\textrm{out}$:
\begin{equation}
\textrm{e}^{C}=\frac{R_\textrm{out}^2}{2\sigma}\left(1-\frac{a_0}{R_\textrm{out}}\right)\textrm{exp}\left(-\frac{R_\textrm{out}^2}{2\sigma\eta}\right)\,.
\end{equation}
Therefore, the metric function $A_2(r_2)$ can be written as
\begin{equation}
A_2\left(r_2\right)=\left(1-\frac{a_0}{R_\textrm{out}}\right)\frac{R_\textrm{out}^2}{r_2^2}\,\textrm{exp}\left({\frac{r_2^2-R_\textrm{out}^2}{2\sigma\eta}}\right)\,.\label{metricint2}
\end{equation}
This exponentially large redshift, $-g_{tt}\sim\textrm{exp}[-(R_\textrm{out}^2-r_2^2)/2\sigma\eta]$, freezes almost entirely the local time at a deep point $r\ll R_\textrm{out}$ as seen by a distant observer within the timescale $\Delta t < e^{R_\textrm{out}^2/\sigma}$.\footnote{This is different from the model of Refs.~\cite{Arrechea:2023hmo,Arrechea:2023oax} in which the redshift inside the object remains moderate, even if the object can also be arbitrarily compact.} However, such a long timescale should not be physical, since the typical timescale of black holes is at most the evaporation timescale $\sim a_0^3/nl_p^2$. Such a hierarchy in different timescales will provide a natural mechanism to darken the images, as we will show in sec.~\ref{sec:image2}.

At this point, the relation between $R_\textrm{out}$ and $a_0$ remains undetermined. It is determined by the continuity of the proper acceleration 
\begin{equation}
\alpha(r)\equiv\frac{\partial_r\ln{\sqrt{-g_{tt}(r)}}}{\sqrt{g_{rr}(r)}}
\end{equation}
at the surface $R_\textrm{out}$: $\alpha_1(R_\textrm{out})=\alpha_2(R_\textrm{out})$ \cite{Yokokura:2022phq}.\footnote{This is required from a thermodynamic equilibrium condition at Hawking temperature \cite{Yokokura:2022phq}, and it is consistent with Israel's junction condition \cite{Poisson_book}.}
This gives the following algebraic relation 
\begin{equation}
a_0^2R_\textrm{out}\sigma\eta^2-2\left(R_\textrm{out}-a_0\right)\left(R_\textrm{out}^2-2\eta\sigma\right)^2=0\,.\label{algebraicRa0}
\end{equation}
Solving this, one can obtain the analytic expression of $a_0=a_0(R_\textrm{out})$ as
\begin{equation}
a_0=a_0(R_\textrm{out})=\frac{-W^2+\sqrt{W^4+2R_\textrm{out}^2W^2\sigma\eta^2}}{R_\textrm{out}\sigma\eta^2}\,,\label{Rexp}
\end{equation}
with $W\equiv R_\textrm{out}^2-2\eta\sigma$. For $\sigma\ll R_\textrm{out}^2$, one can expand Eq.~\eqref{Rexp} and obtain the approximated relation \cite{Yokokura:2022phq}:
\begin{equation}
R_\textrm{out}\approx a_0+\frac{\sigma\eta^2}{2a_0}\,.\label{Raapp}
\end{equation}

It is convenient for the following analysis to consider, instead of $\sigma$, another dimensionless parameter 
\begin{equation}\label{k_def}
k\equiv \frac{\sigma\eta^2}{a_0^2}\,.
\end{equation}
From the relation \eqref{sigma}, we have $k\sim n l_p^2/a_0^2$. The parameter $k$ can roughly quantify the difference in size between the QHCO surface and $a_0$ because, from Eq.~\eqref{Raapp}, we get $R_\textrm{out}/a_0-1\approx k/2$ when $k\ll1$. We here discuss the typical value of $k$. For a QHCO about 10 times the solar mass, we have $a_0\sim 3\times 10^4 m$, 
and thus we obtain $k\sim  10^{-79}n$, where we use $l_p\sim 10^{-35} m$. 
If we assume the Standard Model at the near-Planckian scale, the number of degrees of freedom in the theory is $n\sim 10^2$. Then, we get $k\sim 10^{-77}$, which is extremely small.

From this estimate, one might think that the quantum effects are just too small to generate any significant horizon-scale effects. However, this is not the case because the interior structure is essentially independent of the total energy $a_0/2G$; the proper length between $a_0$ and $R_\textrm{out}$ is estimated from Eqs.~\eqref{dense_metric},\eqref{eta}, \eqref{sigma}, and \eqref{Raapp} as $\sqrt{g_{rr}^{(2)}}\frac{\sigma\eta^2}{2a_0}\sim \sqrt{n}l_p$, and the scale of the interior curvatures is $\sim1/nl_p^2$ as a result of non-perturbative quantum effects like the large pressure $\langle T^\theta{}_\theta\rangle\sim1/Gnl_p^2$. In the following sections, to demonstrate the phenomenological effects of $k$ in a straightforward manner, the analysis will be performed using a modestly large value of $k$. Therefore, for the rest of this paper, we will use the exact expression given by Eq.~\eqref{Rexp} whenever we calculate the QHCO surface $R_\textrm{out}$.

Finally, the central small region of size $\sim\sqrt{n}l_p$ is flat and the inner surface can be chosen at, say \cite{Kawai:2020rmt,Yokokura:2023wxp},
\begin{equation}
    R_\textrm{core}=\sqrt{2\sigma}\,.
\end{equation}
More explicitly, after connecting to the metric \eqref{dense_metric} in which $A_2(r_2)$ is given by Eq.~\eqref{metricint2}, 
the metric of the central core region is given by 
\begin{align}
A_3\left(r_3\right)&=\left(1-\frac{a_0}{R_\textrm{out}}\right)\frac{R_\textrm{out}^2}{2\sigma}\,\textrm{exp}\left({\frac{2\sigma-R_\textrm{out}^2}{2\sigma\eta}}\right)=\textrm{const.}\,,\nonumber\\
B_3\left(r_3\right)&=1\,,\label{metricint3}
\end{align}
which can be seen to be flat by redefining the time coordinate.

\section{Photon trajectories}\label{sec.photoneq}
To construct the QHCO image, we first investigate how light rays propagate in the QHCO spacetime. Because there is no event horizon, light rays that enter the QHCO surface $R_\textrm{out}$ can in principle escape to the exterior again. The construction of the images has to take into account those light ray trajectories as well. The photon trajectories outside the QHCO surface are completely determined by the geodesic equations of the Schwarzschild spacetime which have been well understood. In this section, therefore, we will mainly focus on the photon geodesics that can enter the surface of QHCO.

The whole spacetime is static and spherically symmetric. Therefore, one can define two constants of motion for a geodesic $x^\mu(\lambda)$ in each spacetime region:
\begin{equation}
A_{i}\dot{t}_i=E_i\,,\qquad r_i^2\dot{\varphi}_i=L_i\,,\label{EL}
\end{equation}
where the dot represents the derivative with respect to the affine parameter $\lambda_i$, and the constants of motion $E_i$ and $L_i$ are the energy and the angular momentum of the geodesic. For the equatorial motion $\theta=\pi/2$, the null constraint condition $g_{\mu\nu}\dot x^\mu \dot x^\nu=0$ gives
\begin{equation}
-A_i\dot{t}_i^2+B_i\dot{r}_i^2+r_i^2\dot{\varphi}_i^2=0\,,
\end{equation}
which, using Eq.~\eqref{EL}, can be rewritten as
\begin{equation}
A_i B_i\dot{r}_i^2+\frac{A_i L_i^2}{r_i^2}=E_i^2\,.\label{constraint}
\end{equation}

Since we focus on the photon trajectories that may cross the two surfaces, we have to take into account the boundary conditions of the photon 4-momenta at the surfaces. Here, considering the geometry near a timelike hypersurface $\bold\Sigma$, one can introduce a Gaussian normal coordinate system \cite{Sakai:2014pga}
\begin{equation}
ds^2=d\tilde{n}^2-\alpha_\pm\left(\tilde{n},\tau\right)^2d\tau^2+r_\pm\left(\tilde{n},\tau\right)^2d\Omega_2^2\,,\label{Gaussiannormal}
\end{equation}
where the subscripts $\pm$ denote the opposite sides of $\bold\Sigma$; the surface corresponds to $\tilde{n}=0$;  $\alpha$ is normalized such that $\alpha_\pm(0,\tau)=1$; and the areal radius of $\bold\Sigma$ is represented by $r_\pm(0,\tau)$. Now, applying this coordinate system \eqref{Gaussiannormal} to the QHCO surface, the continuity of the $d\tau^2$ and the $d\Omega_2^2$ terms gives
\begin{equation}
A_1(R_\textrm{out})\dot{t}_1^2=A_2(R_\textrm{out})\dot{t}_2^2\,,\quad r_1^2\dot\varphi_1^2=r_2^2\dot\varphi_2^2\,,
\end{equation}
respectively, with $r_1=r_2=R_\textrm{out}$. A similar procedure applies to the core surface at $r_2=r_3=R_\textrm{core}$. Because the metric functions $A_i$ and $r_i$ are continuous at the two surfaces, one eventually obtains
\begin{equation}
\dot{t}_1=\dot{t}_2=\dot{t}_3\,,\qquad \dot\varphi_1=\dot\varphi_2=\dot\varphi_3\,.\label{dot123}
\end{equation}
Combining Eqs.~\eqref{EL} and \eqref{dot123}, one gets
\begin{equation}
E_1=E_2=E_3\,,\quad L_1=L_2=L_3\,.
\label{conservation}
\end{equation}
Therefore, the energy and angular momentum are conserved along the geodesic in the whole space, and then we can freely drop the subscripts $i$ for these constants of motion.

Then, evaluating Eq.~\eqref{constraint} at the surfaces $R_\textrm{out}$ and $R_\textrm{core}$ with the conservation laws \eqref{conservation} and the continuity of $A_i$ and $r_i$, one gets the following relations:
\begin{equation}
\sqrt{B_1(R_\textrm{out})}\dot{r}_1=\sqrt{B_2(R_\textrm{out})}\dot{r}_2\,,
\end{equation}
and
\begin{equation}
\dot{r}_2=\dot{r}_3\,,\label{radialcore}
\end{equation}
where we have explicitly used $B_2=B_3$ at the core surface $R_\textrm{core}$. 

With the relation of the 4-momenta in different spacetime regions, we define the following unified variables
\begin{align}
\textrm{Exterior Schwarzschild}:&\quad r=r_1\quad\textrm{and}\quad\lambda=\lambda_1\,,\nonumber\\
\textrm{Dense region}:&\quad r=r_2\quad\textrm{and}\quad\lambda=\sqrt{\frac{B_1(R_\textrm{out})}{B_2(R_\textrm{out})}}\lambda_2\,,\nonumber\\
\textrm{Central flat core:}&\quad r=r_3\quad\textrm{and}\quad\lambda=\sqrt{\frac{B_1(R_\textrm{out})}{B_2(R_\textrm{out})}}\lambda_3\,,\nonumber\\
\end{align}
The constraint equation \eqref{constraint} can be written as
\begin{equation}
\frac{C_i(r)}{L^2}\left(\frac{dr}{d\lambda}\right)^2+V(r)=\frac{1}{b^2}\,,\label{radialformula}
\end{equation}
where
\begin{align}
C_1(r)=1\,,&\quad C_2(r)=\frac{B_1(R_\textrm{out})}{B_2(R_\textrm{out})}A_2(r)B_2(r)\,,\nonumber\\C_3(r)&=\frac{B_1(R_\textrm{out})}{B_2(R_\textrm{out})}A_3(r)\,,
\end{align}
$b\equiv L/E$ is the impact parameter of photon trajectories, and the effective potential $V(r)$ reads
\begin{align}
\label{Vr}
V\left(r\ge R_\textrm{out}\right)&=\frac{A_1(r)}{r^2}\,,\nonumber\\
V\left(R_\textrm{core}\le r<R_\textrm{out}\right)&=\frac{A_2(r)}{r^2}\,,\nonumber\\
V\left(r<R_\textrm{core}\right)&=\frac{A_3(r)}{r^2}\,.
\end{align}

\begin{figure}[h]
  \centering
 \includegraphics[scale=0.48]{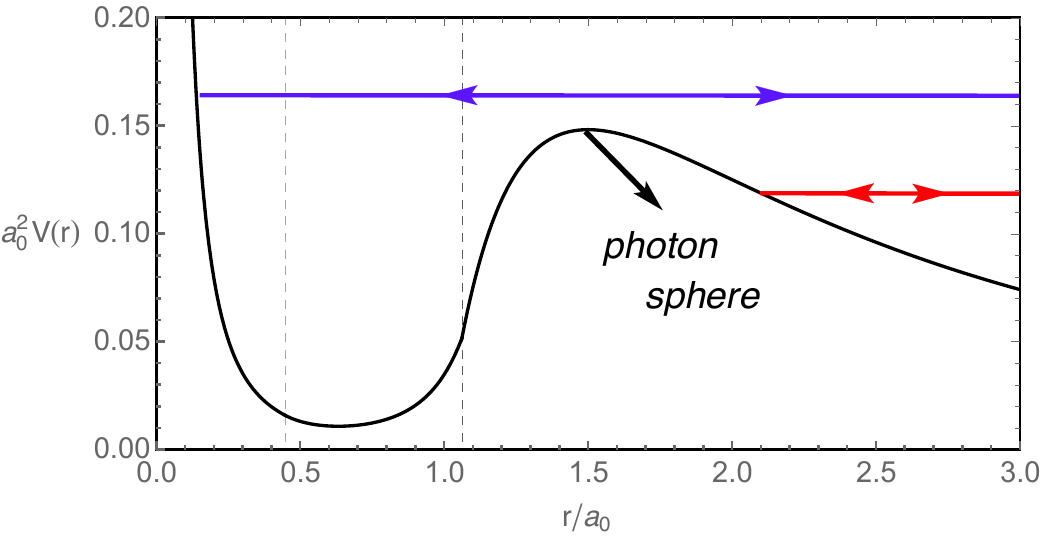}
\\
\caption{The effective potential \eqref{Vr}, $V(r)$, with $k=1/10$ and $\eta=1$. The two vertical dashed lines correspond to the outer surface $R_\textrm{out}$ (right) and the core surface $R_\textrm{core}$ (left). Photon trajectories with large impact parameters (red) acquire turning points at $r=r_\textrm{tp}$ outside the photon sphere $r_p$. On the other hand, trajectories with small impact parameters (blue) enter the QHCO and have turning points at $r=r_\textrm{tp}$ inside. 
}
\label{fig:schematicpo} 
\end{figure}

In Fig.~\ref{fig:schematicpo}, we show the effective potential $V(r)$ with the parameters $\eta=1$ and $k=1/10$ as an illustration. 
The vertical dashed lines indicate the outer surface $R_\textrm{out}$ (right) and the core surface $R_\textrm{core}$ (left). Outside $R_\textrm{out}$, the effective potential is given by the Schwarzschild one and has a peak at $r_p=3a_0/2$. This peak corresponds to the photon sphere on which photons can undergo unstable circular motions. The photon sphere and its vicinity are characterized by highly lensed photon trajectories, as we have discussed in the Introduction. Inside the outer surface, the effective potential has a local minimum at $r_\textrm{min}=2\sqrt{\sigma\eta}$, which can be obtained by solving $\partial_r{V(r)}=0$ inside $R_\textrm{out}$.{\footnote{The minimum of the effective potential could correspond to a stable photon sphere which may lead to instability issues \cite{Keir:2014oka,Cardoso:2014sna,Cunha:2022gde,Carballo-Rubio:2023mvr}, although trajectories around $r=r_\textrm{min}$ are not physically relevant in a timescale $\Delta t\sim a_0^3/\sigma$ due to the large redshift, as will be discussed in sec.~\ref{sec:image2}.}} After crossing the minimum, the effective potential increases as one moves inward, forming a centrifugal barrier inside the central flat core.

The expression of Eq.~\eqref{radialformula} is particularly useful in identifying the radial turning point for each photon trajectory. The turning point $r_\textrm{tp}$ of a trajectory with impact parameter $b$ can be calculated simply by solving $V(r_\textrm{tp})=1/b^2$. According to Fig.~\ref{fig:schematicpo}, photon trajectories with $b^2>1/V(r_p)$ have turning points outside the unstable photon sphere $r_p$, and hence do not enter the QHCO. These trajectories are illustrated by the red line in Fig.~\ref{fig:schematicpo}. On the other hand, photon trajectories with $b^2<1/V(r_p)$, which are illustrated by the blue line, enter the object, go all the way down into the central core, reach the turning point inside the core, and then radially return. Note that this is the motion as seen in terms of the affine parameter $\lambda$. 

\begin{figure}[h]
  \centering
 \includegraphics[scale=0.65]{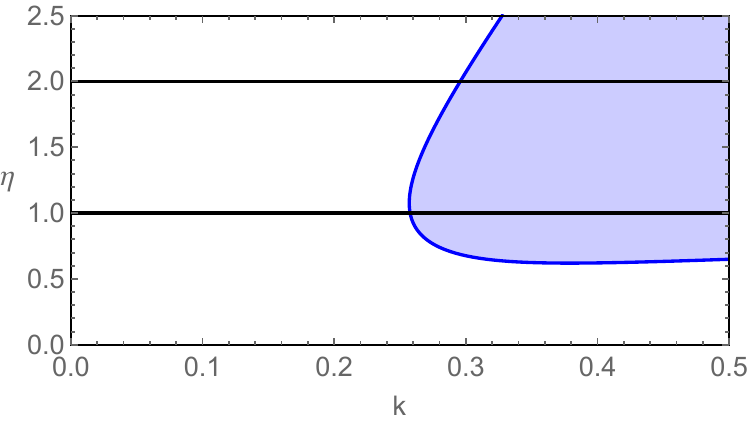}
\caption{The parameter space (shaded) in which there exist infalling photon trajectories that do not reach the central core. The two black lines indicate $\eta=1$ and $\eta=2$.}
\label{fig:entercore} 
\end{figure}

Consider now the photons that enter the outer surface $R_\textrm{out}$ (called \textit{infalling photons}). Whether all of them can further reach the central core depends on the choices of the parameters $(k,\eta)$ (in Fig.~\ref{fig:schematicpo}, the parameters are chosen so that this is the case). Essentially, if one wants some infalling photons to return within the dense region, the effective potential at the unstable photon sphere, $r_p$, has to be lower than that at the core surface $R_\textrm{core}$: $V(r_p)<V(R_\textrm{core})$. In Fig.~\ref{fig:entercore}, we show such a region of the parameter space (shaded region). 
Therefore, in the shaded region, there exist photons that enter the dense region but not the central core. Equivalently, in the unshaded region, as long as photons enter $R_\textrm{out}$, they always have radial turning points inside the central core. Especially, in the parameter region of physical interest, i.e., $1\le\eta<2$ and $k\ll1$, all the photons that enter the QHCO pierce the central flat core and come back to the outside (as seen in terms of the affine parameter). 

\begin{figure}[h]
  \centering
 \includegraphics[scale=0.45]{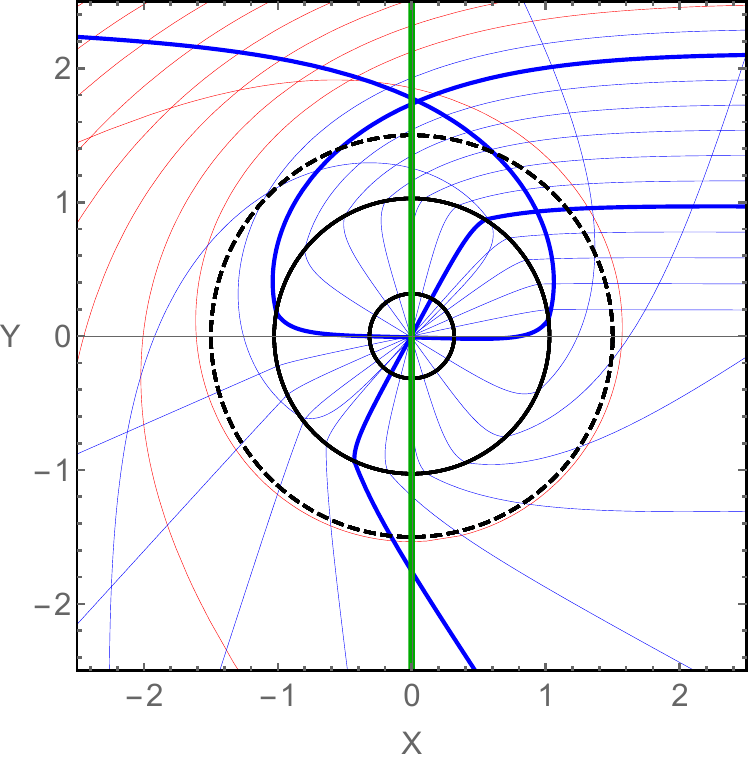}
 \includegraphics[scale=0.45]{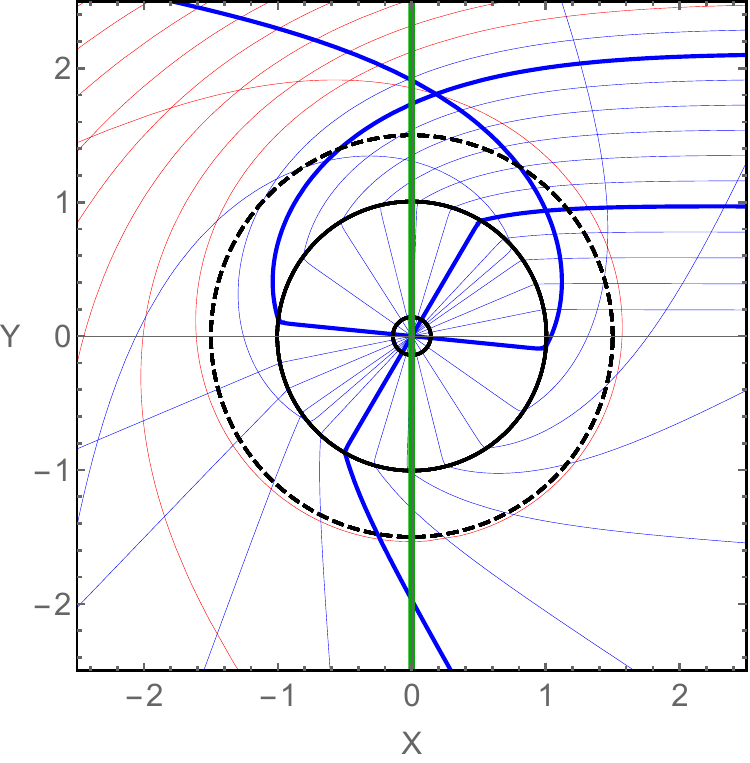}
\caption{Photon trajectories parameterized by the affine parameter in the QHCOs with $\eta=1$, $k=1/20$ (upper) and $k=1/100$ (lower), respectively. The observer is located at $(X_o,Y_o)=(100,0)$. The dashed circle represents the unstable photon sphere $r_p$. The outer surface $R_\textrm{out}$ and the core surface $R_\textrm{core}$ are shown by the two black solid circles. The green lines correspond to the thin disk of light sources (see sec.~\ref{sec:image}). Two specific infalling trajectories are highlighted by the thick blue curves.}
\label{fig:trajectory} 
\end{figure}

Let us plot the photon trajectories, which can be done by integrating the equations of motion \eqref{EL} and \eqref{radialformula}. 
Since they are all planar motions, one can demonstrate the trajectories in the Cartesian coordinates $(X,Y)=(r\cos\varphi,r\sin\varphi)/a_0$. The results are shown in Fig.~\ref{fig:trajectory}, where we fix $\eta=1$ and consider two different QHCO models, one with $k=1/20$ (upper) and the other with $k=1/100$ (lower). 
In this figure, all trajectories are connected to an observer who is located at $(X_o,Y_o)=(100,0)$. The dashed circle represents the unstable photon sphere which has radius $r_p=3a_0/2$. The outer and inner solid circles stand for the outer and core surfaces, respectively. The red and blue curves correspond to the trajectories that have radial turning points $r_\textrm{tp}$ outside and inside the QHCO, respectively. As we have mentioned, the photon trajectories that have impact parameters smaller than that of the unstable photon sphere can all enter the central core region. In addition, unlike the case of classical black holes in which photons entering the event horizon never escape, the infalling photons in the QHCO model can escape and may be observed. When illuminated by the emission from a thin disk of light sources discussed in sec.~\ref{sec:image} (green lines), these infalling photons may pierce through the emission both before entering and after leaving QHCOs, as demonstrated by the highlighted blue trajectories in Fig.~\ref{fig:trajectory}. These additional piercings may give rise to extra intensity on the images. We will discuss in more detail the images of QHCO models in sec.~\ref{sec:image}.

We here introduce two additional crucial quantities relevant to constructing the images of QHCOs in the following sections: the elapsed coordinate time $\Delta t_\textrm{ec}$ associated with a trajectory during its propagation, and the strongest redshift $Q$ experienced along the geodesic. The consideration of the elapsed coordinate time $\Delta t_\textrm{ec}$ is more physical as compared with the analysis in terms of the affine parameter because the latter does not include redshift effects properly. The strong redshift inside QHCOs, on the other hand, is quantified by $Q$.
\begin{figure*}
  \centering
\subfigure[]{\includegraphics[scale=0.595]{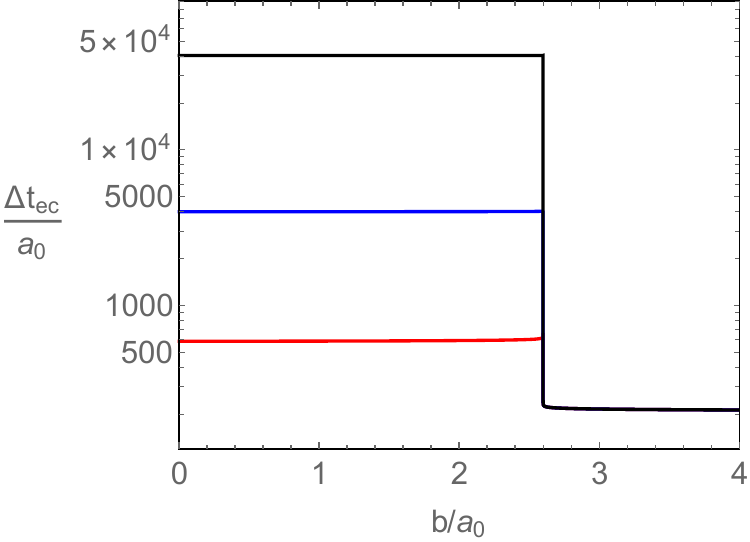}}
 \subfigure[]{\includegraphics[scale=0.59]{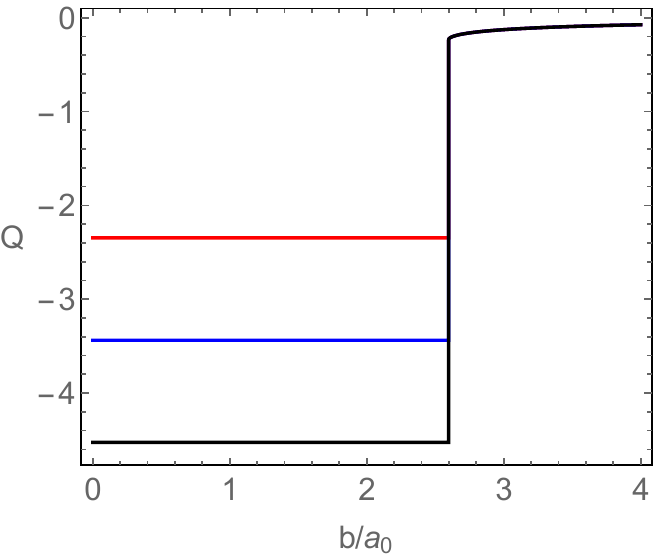}}
\\
\caption{(a) The elapsed coordinate time $\Delta t_\textrm{ec}$ and (b) the strongest redshift $Q\equiv\textrm{min}_{\bold{x}}\left[\log_{10}\sqrt{g_{tt}(\bold{x})/g_{tt}(\bold{x}_o)}\right]$ experienced along the geodesic. We fix $\eta=1$, and vary the values of $k=1/20$ (red), $1/30$ (blue), and $1/40$ (black). The vertical jumps correspond to the impact parameter of the unstable photon sphere.}
\label{fig:potential} 
\end{figure*} 

The elapsed coordinate time $\Delta t_\textrm{ec}$ along a photon propagation  can be formally expressed as
\begin{equation}\label{dt_ec}
\Delta t_\textrm{ec}=\int_{\lambda_{\textrm{initial}}}^{\lambda_\textrm{final}} \dot{t}d\lambda\,,
\end{equation}
where the final value of the affine parameter $\lambda_\textrm{final}$ corresponds to the moment when the photon reaches the observer $(X_o,Y_o)$. The initial point for the trajectory, on the other hand, is assumed to be at $r=100 a_0$.{\footnote{The exact initial point is not relevant because the elapsed time is dominated by the propagation inside QHCOs.}} The results are shown in Fig.~\ref{fig:potential}(a). For photons entering the outer surface $R_\textrm{out}$, $\Delta t_\textrm{ec}$ is extremely large due to the exponentially large redshift in the interior metric \eqref{dense_metric}. In particular, for a more realistic case where $k\ll1$, $\Delta t_\textrm{ec}$ is dominated by the propagation in the dense region and can be calculated through the metric \eqref{dense_metric} approximately as
\begin{align}
  \ln{\frac{\Delta t_\textrm{ec}}{a_0}}&\approx\ln{\frac{(\Delta t_\textrm{ec})|_{\textrm{region 2}}}{a_0}}\approx\frac{\eta}{4k}\nonumber\\
  \Rightarrow \Delta t_\textrm{ec}&\sim a_0 e^{a_0^2/nl_p^2}\,.
  \label{exptime} 
\end{align} 
This indicates that at the leading order, the elapsed coordinate time for infalling photons is exponentially large and independent of the impact parameter $b$.

The exponential time dilation given in Eq.~\eqref{exptime} is due to the strong redshift experienced by the photons inside QHCOs. We can quantify the strength of redshifts of trajectories by computing the strongest redshift $Q$ experienced along the geodesic, i.e., $Q\equiv\textrm{min}_{\bold{x}}\left[\log_{10}\sqrt{g_{tt}(\bold{x})/g_{tt}(\bold{x}_o)}\right]$, where $\bold{x}_o=(X_o,Y_o)$ is the position of the observer. The results are shown in Fig.~\ref{fig:potential}(b). Similar to Eq.~\eqref{exptime}, the strongest redshift experienced by photons entering QHCOs is also independent of $b$. This is because the strongest redshift appears at $r=\sqrt{2\sigma\eta}$, which can be derived by solving $\partial_rg_{tt}(r)=0$ in the metric \eqref{dense_metric}, and this is outside the central core $R_\textrm{core}=\sqrt{2\sigma}$ as long as $\eta\ge1$. As we have mentioned, all the trajectories entering the QHCOs with $(k,\eta)$ in Fig.~\ref{fig:potential} reach the central core. Therefore, they all pass through the strongest redshift point during their propagation and share the same $Q$.  

We note here that a similar extreme time dilation experienced by photon trajectories has also been discovered in other models of horizonless compact objects, such as fuzzballs \cite{Bacchini:2021fig} and topological solitons \cite{Heidmann:2022ehn}. In those models, the compact objects consist of a collection of different microstructures that can be motivated by string theory. In those models, the compact objects are no longer spherically symmetric. Furthermore, the Liouville integrability for geodesic dynamics is generically lost. Therefore, in those models, the extreme time dilation happens due to the combination of strong redshifts and the chaotic behaviors of trajectories when propagating through the interior. However, in the QHCO models considered here, the trajectories are not chaotic due to the spacetime symmetry, and the extreme time dilation appears due to the property of strong redshifts of the interior metric \eqref{dense_metric}.

\section{Ideal QHCO images}\label{sec:image}
As discussed above, the absence of the event horizon for the QHCO model allows the existence of photons that have traveled through QHCOs before being observed. It is then interesting to understand how these photons may affect, in principle, the images of QHCOs as seen by a distant observer. In this section, we study a very idealized scenario in which such penetrating photons reach the observer no matter how strongly they are redshifted inside QHCOs. Then, we show such ``ideal" images of QHCOs, which have different image features from those of classical Schwarzschild black holes, although they are not physical in the sense that the effect of a physical timescale is not considered here (see sec.~\ref{sec:image2}).

As a simple model of light sources, we consider a geometrically and optically thin accretion disk and 
construct the intensity profiles of observed photons and the associated images of QHCOs. As the materials on the disk gradually spiral inward due to their interactions with gravity and friction, their temperature increases with the emission of electromagnetic radiation. Therefore, the accretion disk can be treated as a disk-shaped light source around the object. The assumption that the disk is optically thin means that the emitted photons are not absorbed when they pierce through the disk again after emission. This allows each photon trajectory to \textit{effectively collect} additional intensity when crossing the disk several times and consequently form higher-order rings on top of the direct emission in the images. Note that the assumption of a geometrically and optically thin disk significantly simplifies the construction of images, while it can already capture the main image features of a compact object surrounded by accreting matters \cite{Gralla:2019xty}.

Moreover, it is commonly presumed that the inclination angle between the line of sight of observers and the axis of the disk is small in the observation of M87*. This is inferred from the observational orientation of the jets \cite{CraigWalker:2018vam}. Therefore, for simplicity, we assume a face-on orientation of the observer with respect to the disk, i.e., the line of sight of the observer is perpendicular to the disk surface. Taking the configuration shown in Fig.~\ref{fig:trajectory} as an explicit demonstration in which the observer is located at $(X_o,Y_o)=(100,0)$, the disk is then on the $X=0$ plane (green lines).

Let us construct the formula of the observed intensity $I_o$ based on the above assumptions.
First, because there is no absorption, the specific intensities $I_{\nu_e}$ and $I_{\nu_o}$ with frequencies $\nu_e$ and $\nu_o$ in the emission frame and observer's reference frame, are related generically via \cite{Lindquist:1966igj}
\begin{equation}
\frac{I_{\nu_e}}{\nu_e^3}=\frac{I_{\nu_o}}{\nu_o^3}\,.
\label{nu_rel}
\end{equation}
For an asymptotic observer, the photon frequencies at the two frames are related through the redshift factor as $\nu_o/\nu_e=\sqrt{|g_{tt}|}$, which in particular means $d \nu_o/\nu_o=d\nu_e/\nu_e$. Second, for simplicity, we assume that the specific intensity of the disk in the emission frame is monochromatic and only depends on the radial coordinate $r$, i.e., 
\begin{equation}\label{Ie_model}
I_{\nu_e}=I_e(r)\delta(\nu_e-\nu_*)\,.    
\end{equation}
As mentioned above, because there is no absorption, for a trajectory $\mathcal{C}$ reaching the observer, the observed intensity $I_o$ has all the contributions $I_{\nu_o}^{(k)}$ from the photons emitted at each point $r_k$ where the trajectory $\mathcal{C}$ pierces the disk plane i.e., the $X=0$ plane in Fig.~\ref{fig:trajectory}. Thus, the formula of the observed intensity $I_o$ for $\mathcal{C}$ is given by  
\begin{align}\label{formula_Io}
I_o&=\sum_{k\in \mathcal{C}} \int d\nu_o I_{\nu_o}^{(k)}\nonumber\\
   &=\sum_{k\in \mathcal{C}} \int d\nu_e^{(k)} \left(\frac{\nu_o}{\nu_e^{(k)}}\right)^4 I_{\nu_e^{(k)}}^{(k)}\nonumber\\
   &=\sum_{k\in \mathcal{C}}(-g_{tt}(r_k))^2 I_e(r_k)
\end{align}
upon a redefinition of the normalization factor chosen to be the peak value of $I_o$. Here, $\nu_e^{(k)}$ represents the frequency in the emission frame of photons emitted at $r_k$; at the second line, the relations \eqref{nu_rel} and $d\nu_o/\nu_o=d\nu_e^{(k)}/\nu_e^{(k)}$ are applied to each $k$; and at the last line, the specific intensity \eqref{Ie_model} is used for each $k$ together with the relation $\nu_o/\nu_e^{(k)}=\sqrt{-g_{tt}(r_k)}$. 

Since each photon trajectory $\mathcal{C}$ can be labeled by its impact parameter $b$, the observed intensity is a function of $b$, i.e., $I_o=I_o(b)$. Furthermore, it should be emphasized that when calculating the observed intensity of the photon trajectories that enter QHCOs in the idealized scenario, we will record all the piercings at $r_k$ of the trajectory through the disk, including those before entering QHCOs as well as those possibly inside QHCOs. The intensity collected by these additional piercings on the ideal images may give rise to image features that can be used to distinguish from a classical black hole image, as we will demonstrate later.

Regarding the emission profile $I_e(r)$ on the disk, we employ the Gralla-Lupsasca-Marrone (GLM) emission model \cite{Gralla:2020srx} whose intensity profile takes the following form:
\begin{equation}\label{GLM}
I_e(r;\gamma,\mu,\tilde\sigma)=\frac{\textrm{exp}{\left\{-\frac{1}{2}\left[\gamma+\textrm{arcsinh}\left(\frac{r-\mu}{\tilde\sigma}\right)\right]^2\right\}}}{\sqrt{\left(r-\mu\right)^2+\tilde\sigma^2}}\,.
\end{equation}
The free parameters $(\gamma,\mu,\tilde\sigma)$ are constants that control the overall shape of the intensity profile. The parameter $\gamma$ controls the rate of decay of the intensity toward $r\rightarrow\infty$. The parameter $\mu$ horizontally shifts the intensity profile and can be used to adjust the location of the intensity peak. The parameter $\tilde\sigma$ is used to adjust the dilation of the profile. In this paper, we consider three different profiles (see Fig.~\ref{fig:emission}):
\begin{figure}[h]
  \centering
 \includegraphics[scale=0.4]{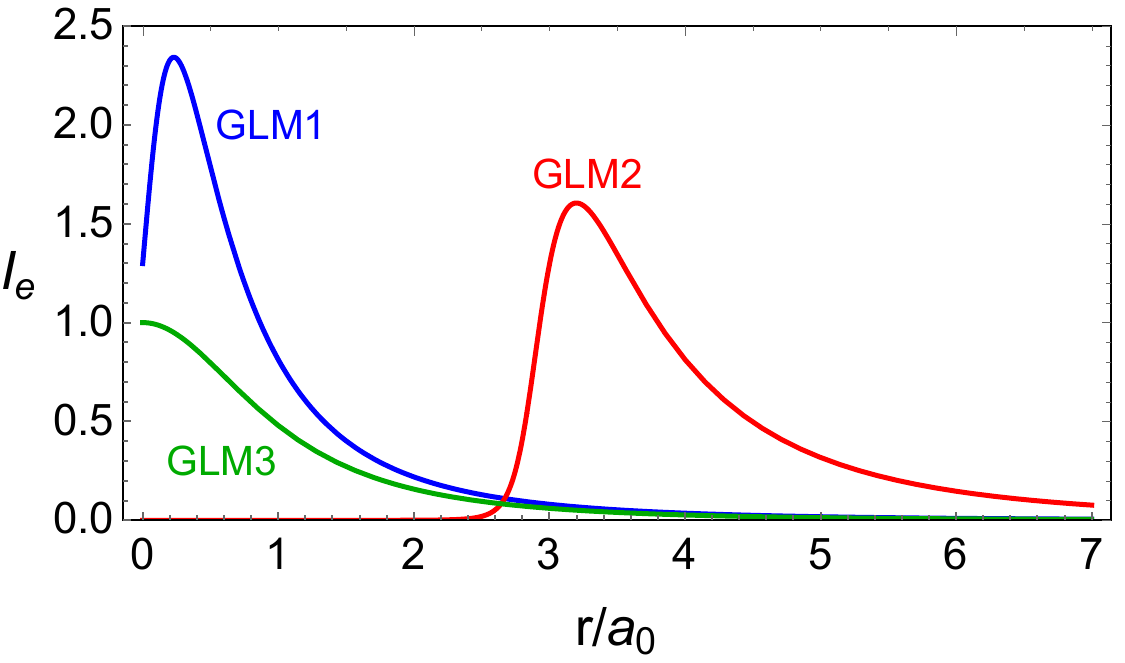}
\caption{The emission profiles $I_e(r)$ of GLM1, GLM2, and GLM3 considered in this work. Note that the photon sphere corresponds to $r_p/a_0=3/2$.}
\label{fig:emission} 
\end{figure}
\begin{itemize}
\item GLM1 profile with $(\gamma,\mu,\tilde\sigma)=\left(-3/2,0,a_0/4\right)$: The intensity profile has a peak slightly outside the origin $r\gtrsim0$, then it decays to zero toward $r\rightarrow\infty$.
\item GLM2 profile with $(\gamma,\mu,\tilde\sigma)=\left(-2,17a_0/6,a_0/4\right)$: The profile has a peak roughly at the innermost stable circular orbit (ISCO) of the Schwarzschild black hole $3a_0$. Below the ISCO, the intensity quickly drops to zero. Also, the intensity decays to zero toward radial infinity.
\item GLM3 profile with $(\gamma,\mu,\tilde\sigma)=\left(0,0,a_0\right)$: The intensity monotonically decays outward, with its peak located at the origin.
\end{itemize}
We would like to mention that when applying the GLM1 and GLM3 profiles to the disk surrounding a classical black hole, the emission extends down to the event horizon. On the other hand, the GLM2 profile may be more astrophysically realistic based on the assumption that the thin disk is composed of a collection of stable circular (Keplerian) orbits of massive particles. In this case, the emission profile naturally drops inside the ISCO on which circular obits become radially unstable. Such a feature of the emission profile is clearly captured by the GLM2 model.

In addition, when constructing the ideal QHCO images, we will collect the intensity of the piercings down to the origin $r=0$. This is based on the implicit assumption that the emission inside QHCOs maintains its disk-shaped structure. Naively, whether such a disk emission exists inside QHCOs should depend on how the accreting materials are captured inside during the formation process of the QHCO and how they interact with the microstructures of the QHCO. However, the intensity collected deep in the QHCOs is significantly suppressed due to the strong redshifts at the emission frame (see the redshift factor in Eq.~\eqref{formula_Io}). Therefore, the intensities contributed by such deep internal piercings can be neglected as compared with those collected from the piercings around and outside QHCOs. Consequently, relaxing the assumptions on the inner edge of the disk emission has only sub-dominant effects on the ideal QHCO images.
 
\begin{figure}[h]
  \centering
 \includegraphics[scale=0.43]{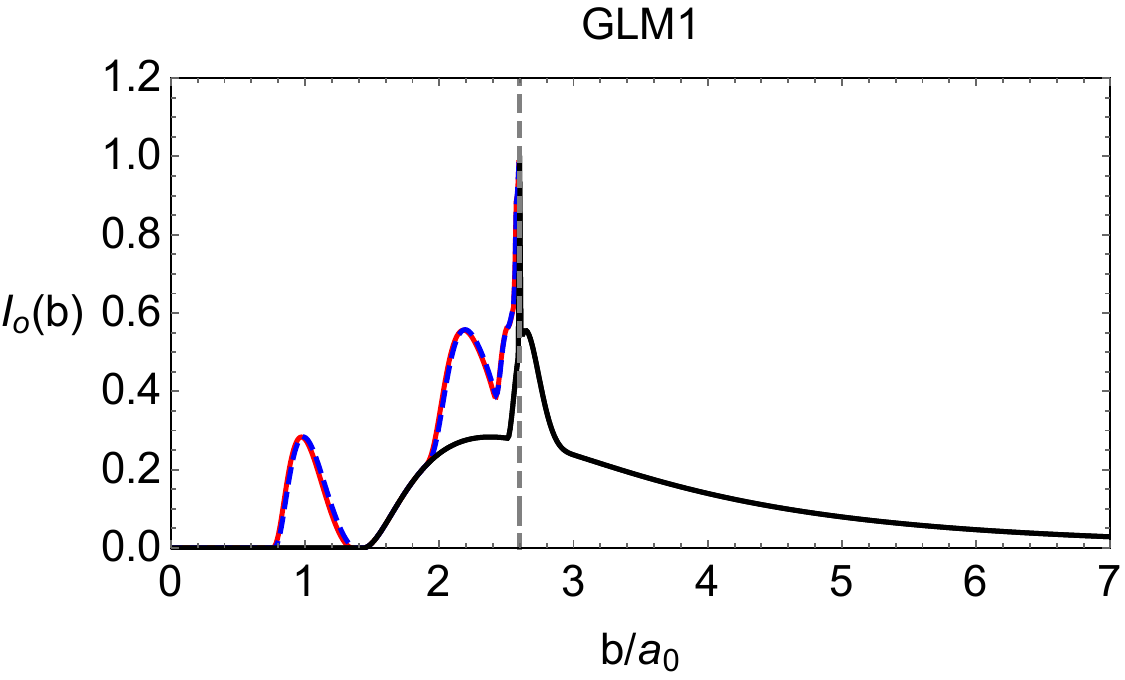}
 \includegraphics[scale=0.22]{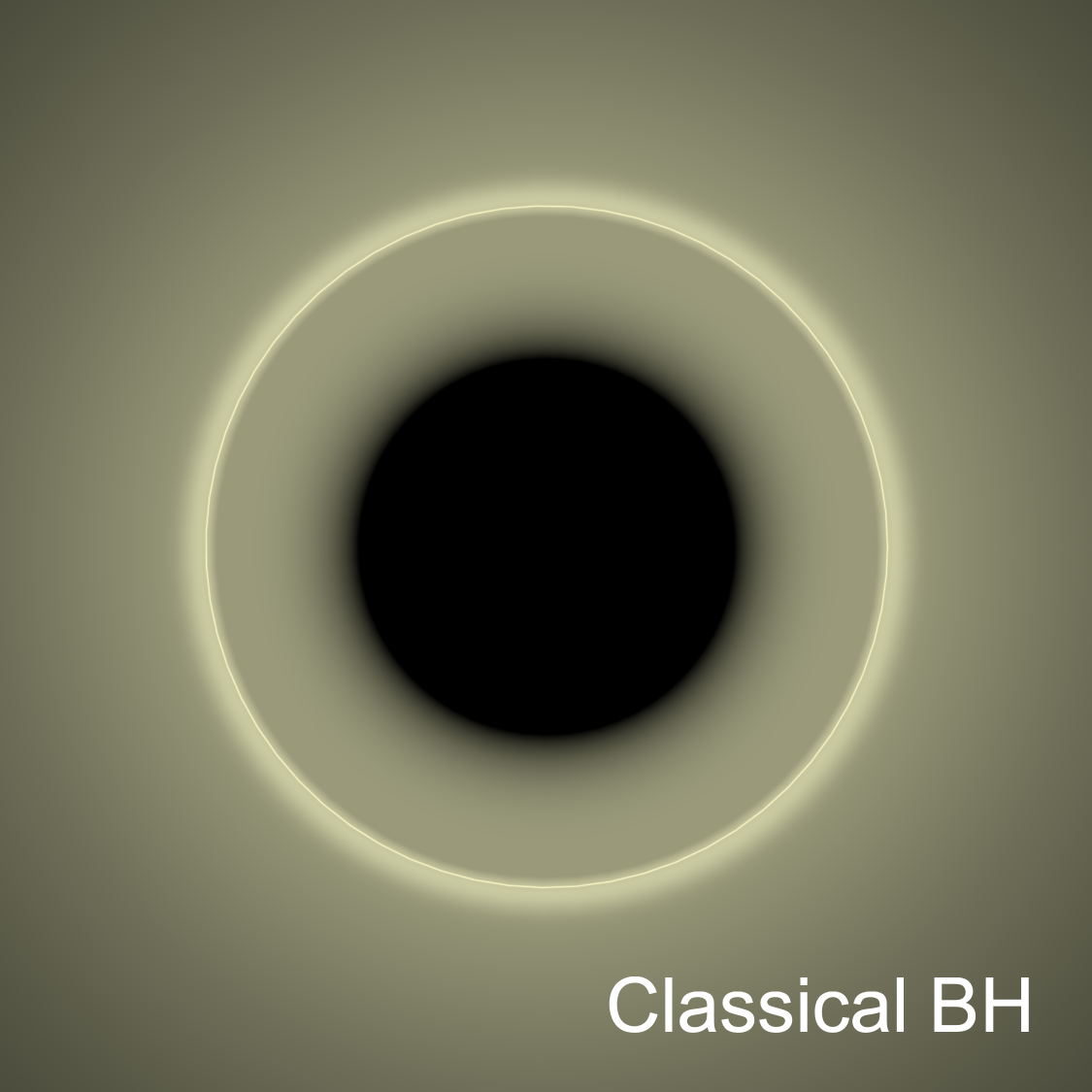}
 \includegraphics[scale=0.22]{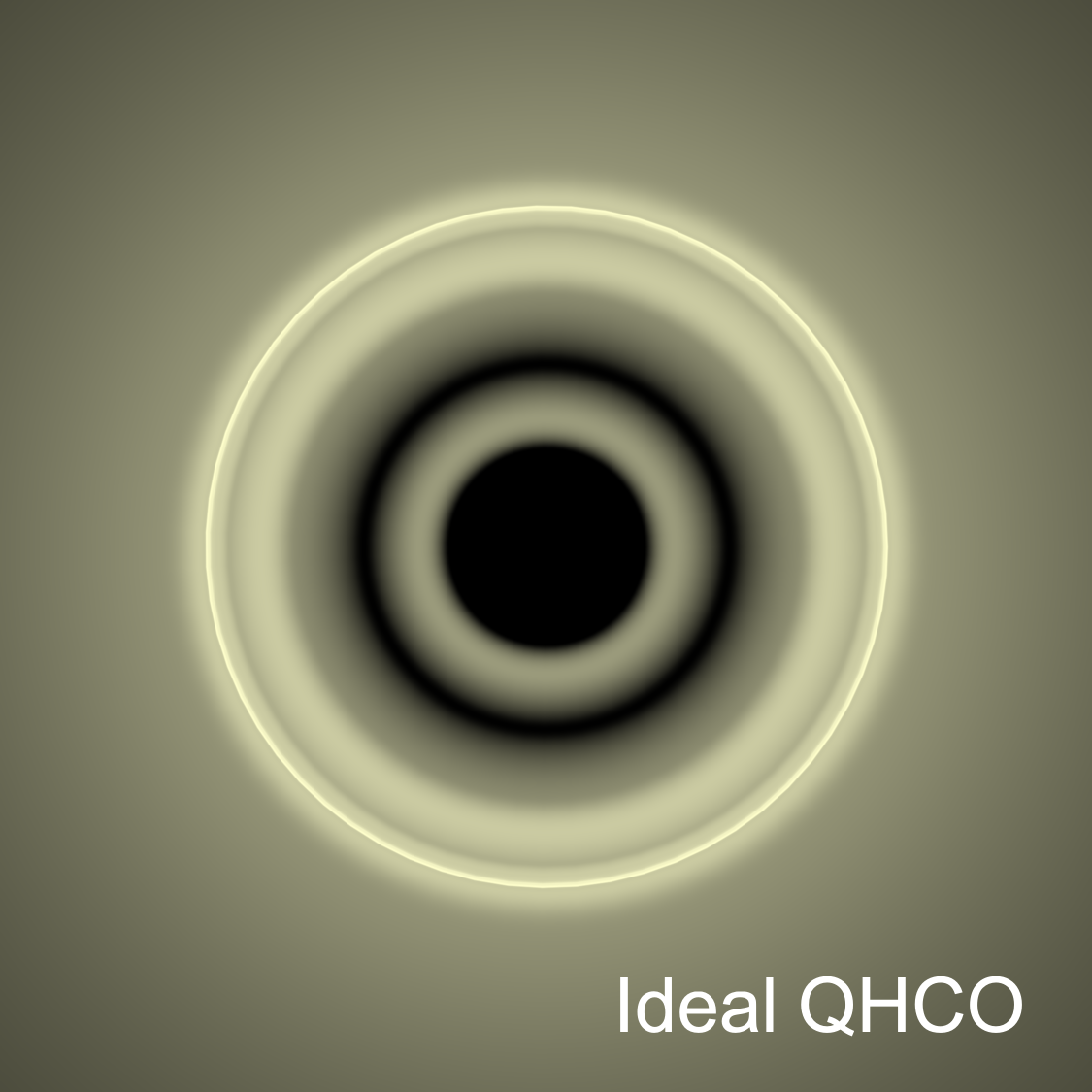}
\caption{The upper panel shows the observed intensity $I_o(b)$ of the classical Schwarzschild black hole (black) and QHCOs with $(k,\eta)=(1/20,1)$ (red) and $(k,\eta)=(1/30,1)$ (blue dashed) surrounded by a thin disk with GLM1 emission profile. In the lower panel, we show the corresponding images of a classical Schwarzschild black hole (left) and the QHCO with $(k,\eta)=(1/20,1)$ (right). Here we consider an idealized scenario in which the photons passing through QHCOs can reach the observer no matter how long their elapsed time $\Delta t_\textrm{ec}$ is.}
\label{fig:image201} 
\end{figure}

\begin{figure}[h]
  \centering
 \includegraphics[scale=0.43]{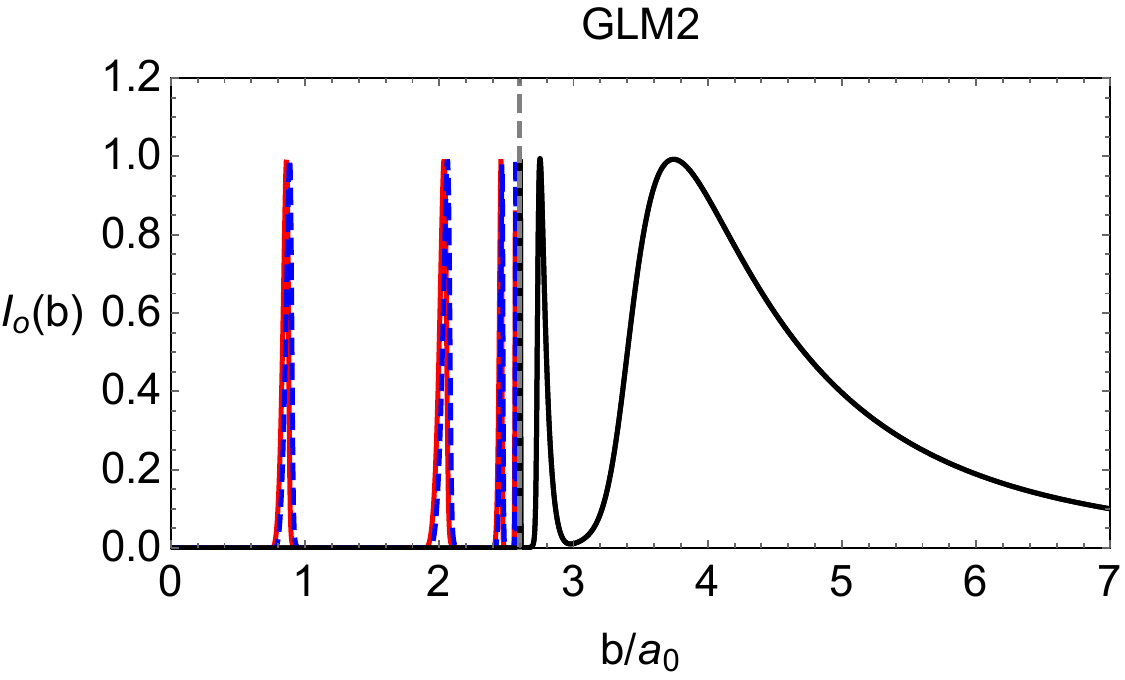}
 \includegraphics[scale=0.22]{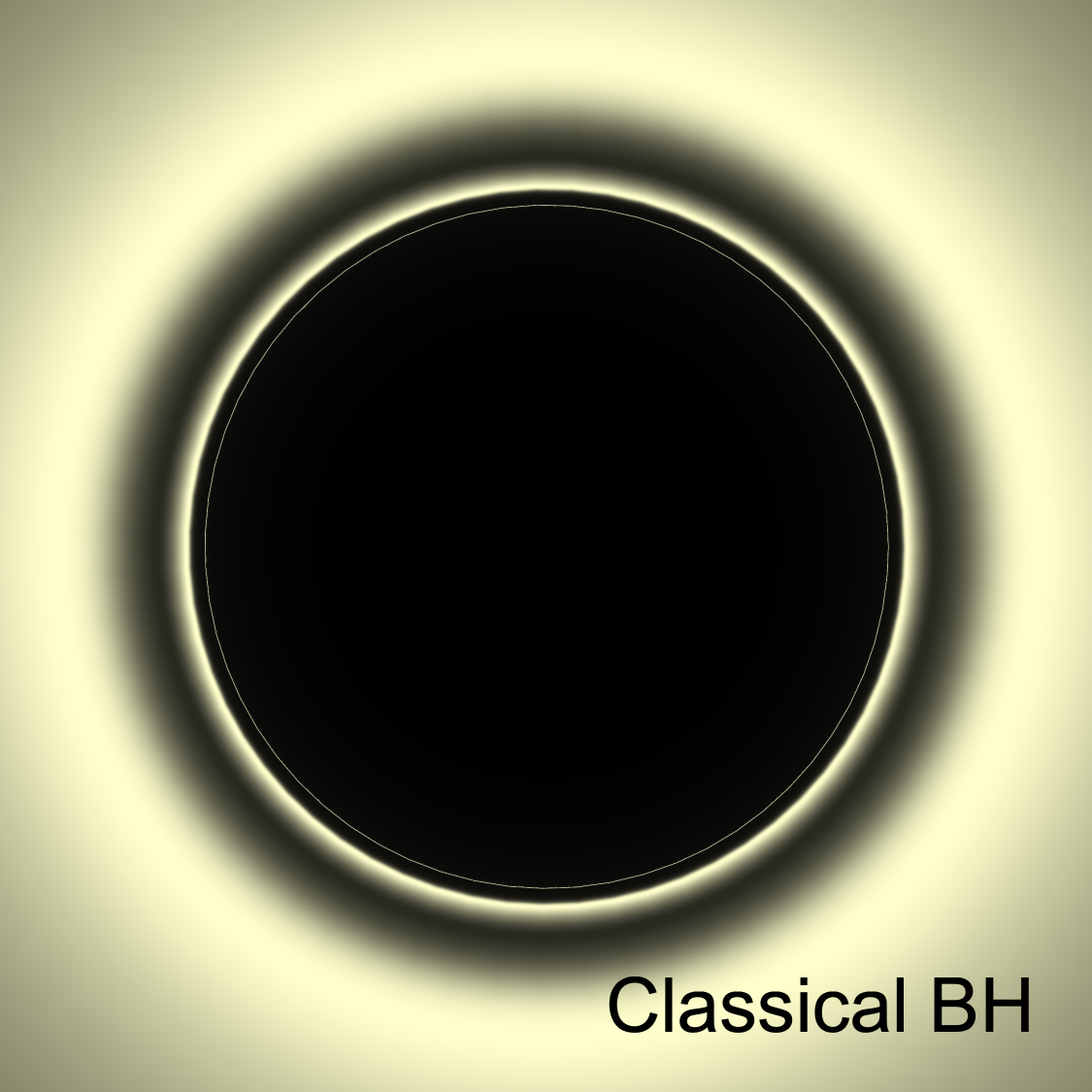}
 \includegraphics[scale=0.22]{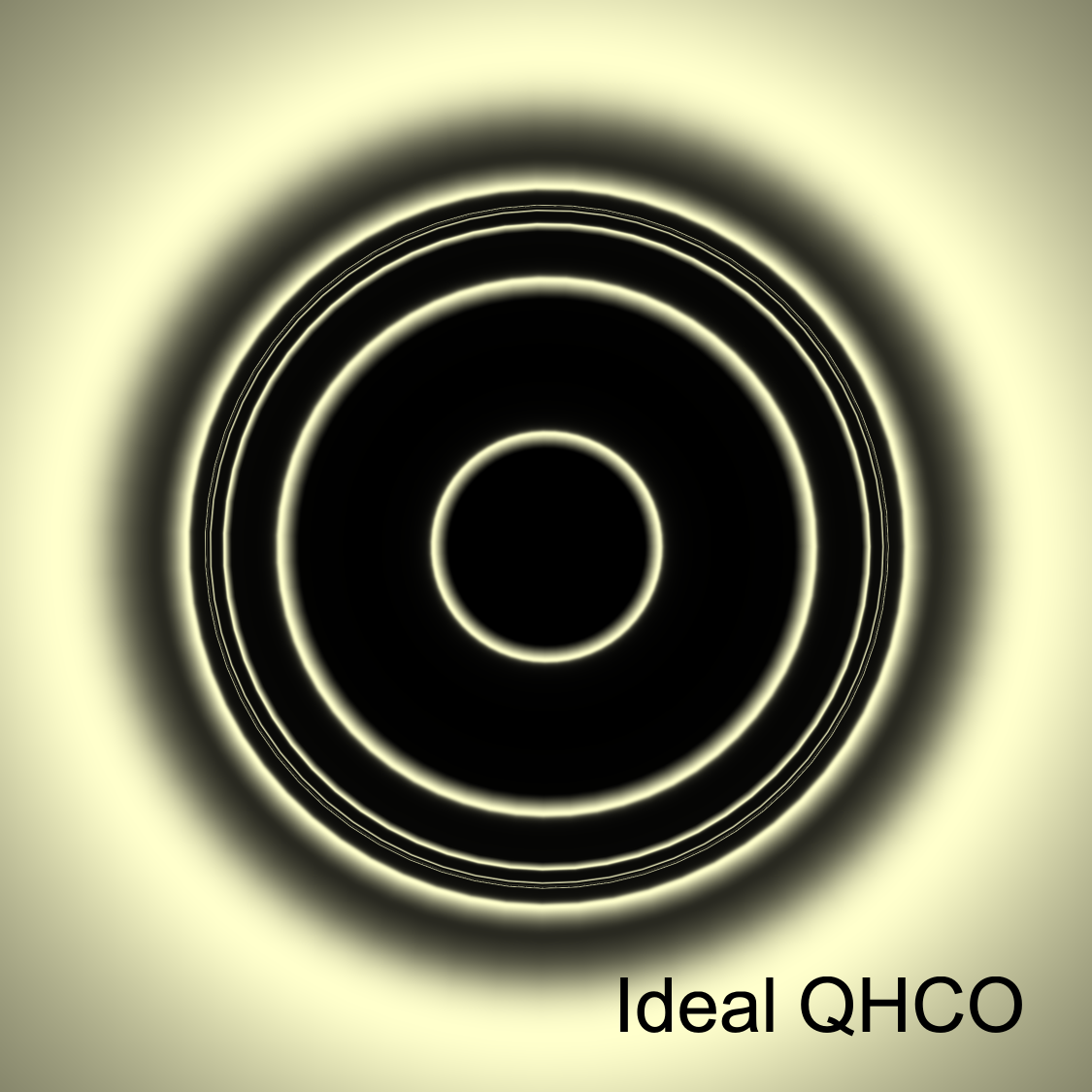}
\caption{The upper panel shows the observed intensity $I_o(b)$ of the classical Schwarzschild black hole (black) and QHCOs with $(k,\eta)=(1/20,1)$ (red) and $(k,\eta)=(1/30,1)$ (blue dashed) surrounded by a thin disk with GLM2 emission profile. In the lower panel, we show the corresponding images of a classical Schwarzschild black hole (left) and the QHCO with $(k,\eta)=(1/20,1)$ (right). Here we consider an idealized scenario in which the photons passing through QHCOs can reach the observer no matter how long their elapsed time $\Delta t_\textrm{ec}$ is.}
\label{fig:image203} 
\end{figure}

Now, we are ready to plot the observed intensity $I_o$ and construct the corresponding ideal QHCO images. We first solve the equations of motion \eqref{EL} and \eqref{radialformula} to obtain the trajectories, from which we identify $r_k$ for each of them; we apply the intensity formula \eqref{formula_Io} with the GLM model \eqref{GLM} and obtain the intensity $I_o(b)$; and then we ``rotate" it around the origin because of the spherical symmetry to get the images. The results are given by Fig.~\ref{fig:image201} and Fig.~\ref{fig:image203} for the GLM1 and GLM2 emission profiles, respectively.\footnote{The intensity profiles and the images of the GLM3 emission are qualitatively very similar to those of GLM1. Therefore, we do not show them here.} Here, the top and lower panels represent, respectively, the observed intensities $I_o(b)$ and the images for the classical black hole and the QHCOs.

First, let us look at the top panel of each figure. The black curve shows the observed intensity $I_o$ for the classical Schwarzschild black hole with the same mass as the QHCOs, and the red and blue curves represent those for the QHCOs with $k=1/20$ and $k=1/30$, respectively, where we fix $\eta=1$.\footnote{Here, in the right part to the critical curve, the three curves are almost identical and therefore appear to be just black.}  Here, the vertical dashed line indicates the impact parameter for the photon sphere called the \textit{critical curve} on the image plane. The wide-spreading smooth lump in $I_o$ comes from the direct emission, which corresponds to the trajectories that only cross the light-source disk once, and its shape highly depends on the emission profiles $I_e(r)$. On the other hand, narrower peaked profiles (called \textit{higher-order images}) exist around the critical curve due to the contribution from the trajectories that rotate around the object many times and cross the disk more than once.

For the GLM1 model (see Figs.~\ref{fig:emission} and \ref{fig:image201}), light emission exists around the photon sphere, and the higher-order images pile up on top of the direct emission around the critical curve. 
For the classical black hole, the black curve in the plot $I_o(b)$ drops to zero at the \textit{inner shadow} that corresponds to the boundary between the bright part and the central dark part in the lower-left image. The inner shadow represents the direct emission from the horizon \cite{Chael:2021rjo}, which only appears when emission at $r=a_0$ exists, as in the GLM1 model. For the QHCOs, on the other hand, the red/blue curve in the plot $I_o(b)$ has additional peaks, which originate from the trajectories that have crossed the disk and collected some intensity before entering the QHCOs (e.g. the thick blue curves highlighted in Fig.~\ref{fig:trajectory}). Because of these additional effects, the observer would see the additional rings inside the bright outer one as shown in the lower-right panel, \textit{if} he/she waited long enough. 

Similarly, we can understand the results for the GLM2 model, which are shown in Fig.~\ref{fig:image203}. In this case, there is no light source around the photon sphere (Fig.~\ref{fig:emission}). The black curve in $I_o(b)$ has still a sharp higher-order image outside the critical curve but becomes zero inside. For the QHCO images, there are additional sharp rings inside the critical curve, which are again generated by photons that have crossed the disk before entering QHCOs.

According to Figs.~\ref{fig:image201} and \ref{fig:image203}, one interesting feature of the ideal QHCO images is that the structure of the additional intensities inside the critical curve is not sensitive to the parameters $(k,\eta)$. This can be seen from the observed intensity profiles of Figs.~\ref{fig:image201} and \ref{fig:image203}, in which the blue and the red curves almost overlap. Does this mean that the inner rings are always visible even if we keep reducing the value of the quantum parameter $k$? To address this question, we have to keep in mind that the images generated in this section are based on the very idealized scenario in which the photons passing through QHCOs can reach the observer no matter how much they are redshifted. In the next section, we will relax this assumption and generate a set of more realistic QHCO images. We will show that, after taking into account the strong redshifts inside QHCOs, the inner rings are hardly observable.

\section{Physical QHCO images}\label{sec:image2}
  
As mentioned just above, the images in Figs.~\ref{fig:image201} and \ref{fig:image203} have been generated based on the idealized assumption that the infalling photons can reach the observer no matter how long their elapsed coordinate time $\Delta t_\textrm{ec}$ \eqref{dt_ec} would be. Such an idealized assumption certainly neglects the possible darkening effects that can be induced by the strong redshifts inside QHCOs and the interaction between photons and the QHCO constituents; indeed, the strong redshifts can delay the propagation of photons, while such interactions may even destroy photons. In this section, we consider these effects and generate more physical images of QHCOs, i.e., \textit{darkened QHCO images}.

To be more concrete about the delays caused by the redshift, we discuss the possible longest timescale for the observation of black-hole images. It should be either the age of the universe or the evaporation timescale of a black hole, $\Delta t_\textrm{eva}\sim a_0^3/nl_p^2$ \cite{Hawking:1975vcx}. For mass $a_0/2G$ larger than the solar mass, the former is shorter than the latter. Noting that QHCOs also evaporate in $\Delta t_\textrm{eva}\sim a_0^3/\sigma$ \cite{Kawai:2013mda,Kawai:2014afa},\footnote{We assume here that QHCOs evaporate completely.} therefore we can consider $\Delta t_\textrm{eva}$ and study the darkening effect here. Then, in principle, one can observe only photon trajectories with their elapsed coordinate time \eqref{dt_ec} not longer than the evaporation time:
\begin{equation}\label{dark_condition}
    \Delta t_\textrm{ec}\lesssim\Delta t_\textrm{eva}\,.
\end{equation}
As we have discussed in sec.~\ref{sec.photoneq}, the elapsed coordinate time of infalling photons is exponentially prolonged due to the strong redshifts inside QHCOs (Fig.~\ref{fig:potential} and Eq.~\eqref{exptime}). As a consequence, the strong redshifts can largely delay the arrival of these infalling photons, effectively darkening the QHCO images inside the critical curve. 

Regarding the internal interactions, it should be noted that a complete consideration of the interaction between internal photon trajectories and the QHCO structure would require detailed analysis for the self-consistent backreaction from the photons and interactions, which could modify the interior metric \eqref{dense_metric} somehow. However, we note here that the parameter $\eta$ phenomenologically quantifies the strength of possible interactions inside QHCOs, as discussed in sec.~\ref{sec.meco}. In the following, we will consider the effect of the internal interaction to be incorporated through $\eta$, although it should be checked eventually whether this treatment is self-consistent in the above sense. 

\begin{figure}[h]
  \centering
 \includegraphics[scale=0.22]{schimage1.pdf}
 \includegraphics[scale=0.22]{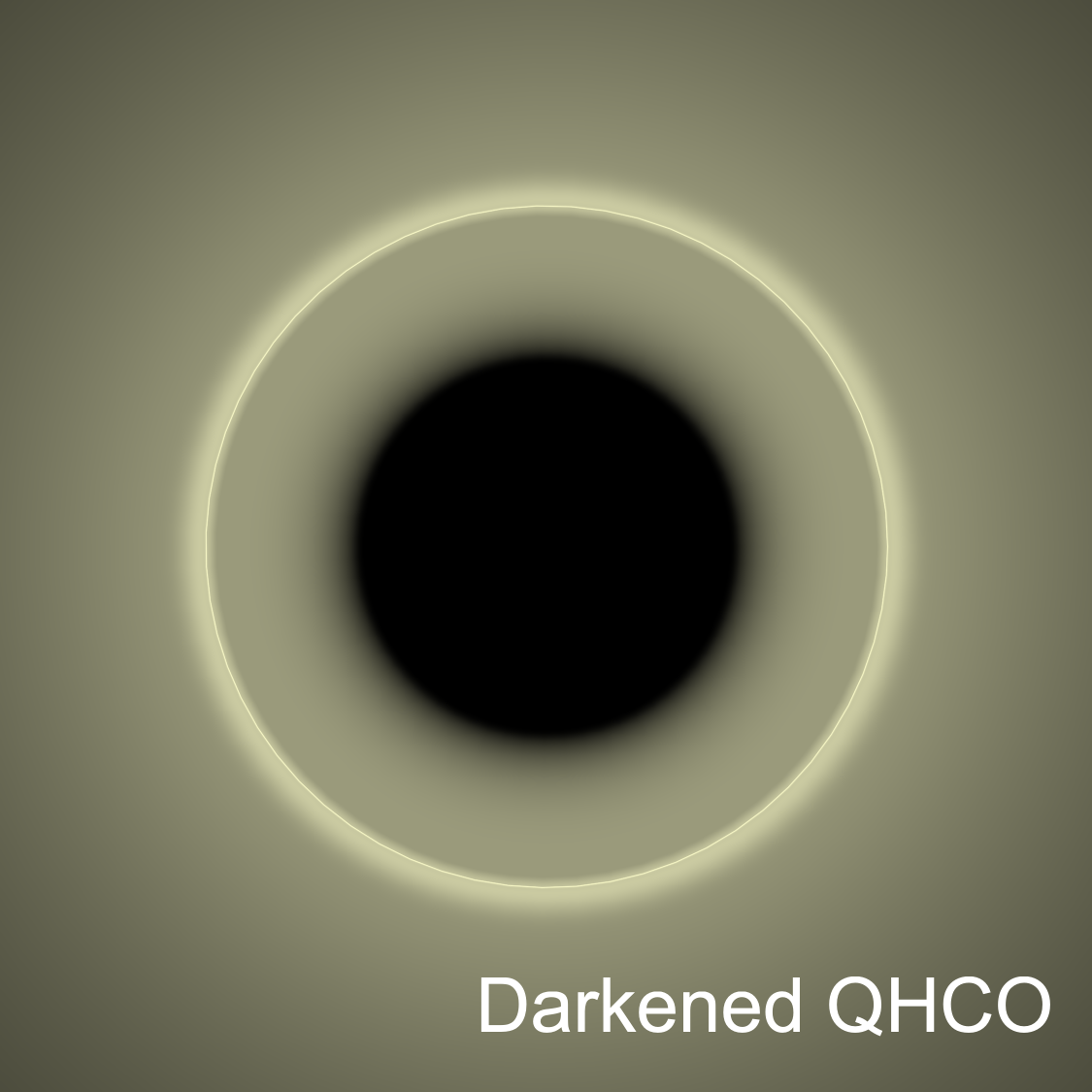}
 \includegraphics[scale=0.22]{schimage3.pdf}
 \includegraphics[scale=0.22]{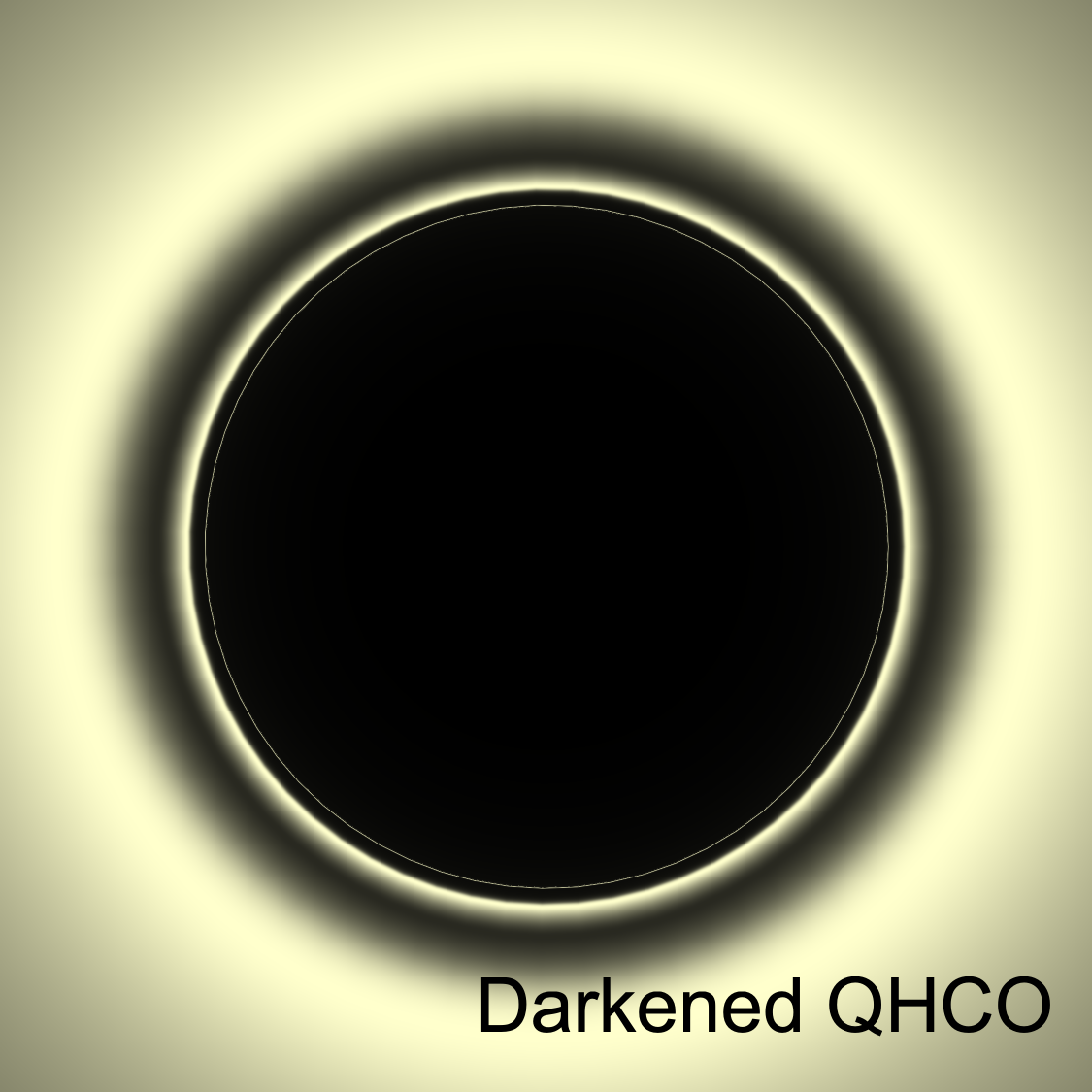}
\caption{The images of the classical Schwarzschild black hole (left) and the darkened images of QHCOs with $k=1/20$ and $\eta=1$ (right). The upper and the lower panels correspond to the GLM1 and GLM2 emission profiles, respectively.}
\label{fig:imagedark} 
\end{figure}

Therefore, to construct the darkened QHCO images, we will follow the argument that only the photons satisfying the condition \eqref{dark_condition} can be observed on the images for various values of $\eta$ in the range of \eqref{eta}. Apparently, implementing this can directly include the aforementioned darkening effects caused by redshifts and internal interactions. Note that a larger value of $\eta$ stands for stronger interactions. Thus, we can gather only the trajectories satisfying the condition \eqref{dark_condition} to generate the darkened QHCO images as in the right panel of Fig.~\ref{fig:imagedark}. Here, the upper-right panel is for the GLM1 and the lower-right one is for the GLM2, where we fix $\eta=1$. 

As shown in Fig.~\ref{fig:imagedark}, basically, the darkened images are very similar to the images of the classical Schwarzschild black hole. This can be understood as follows. We first note that the trajectories through the deep interior of the QHCO generate the inner rings in Figs.~\ref{fig:image201} and \ref{fig:image203}, which are absent in the image of the classical black hole. The elapsed time of such trajectories is exponentially prolonged (as discussed in sec.~\ref{sec.photoneq}), which is much longer than $\Delta t_\textrm{eva}$, and they are cut off in gathering the trajectories satisfying \eqref{dark_condition}. Therefore, the only photon trajectories that can contribute additionally to the inner part of the darkened images are those that are emitted at the shallow part of the QHCOs, i.e., from the outer surface $R_\textrm{out}$ to a depth of $\Delta r \sim \sigma/a_0$ at most, where the strong redshift in the metric \eqref{dense_metric} has little effect. Such photons are very few. As a result, the additional inner rings in Figs.~\ref{fig:image201} and \ref{fig:image203} disappear, and the darkened images in Fig.~\ref{fig:imagedark} are very close to those of the classical cases.

Although it is impossible to distinguish these two sets of images by the naked eye, some subtle differences will appear when there are light emissions near the ``would-be horizon" at $r=a_0$, e.g. the GLM1 and GLM3 emission models, since the photons emitted there may be observable. Because the direct emission of the event horizon (or the would-be horizon for QHCOs) corresponds to the inner shadow in the images, we can expect that some subtle differences between the darkened QHCO images and the Schwarzschild images would appear near their inner shadows, and they would originate from the photons emitted in the near-surface region with width $\Delta r \sim \sigma/a_0$. 

\begin{figure}[h]
  \centering
 \includegraphics[scale=0.43]{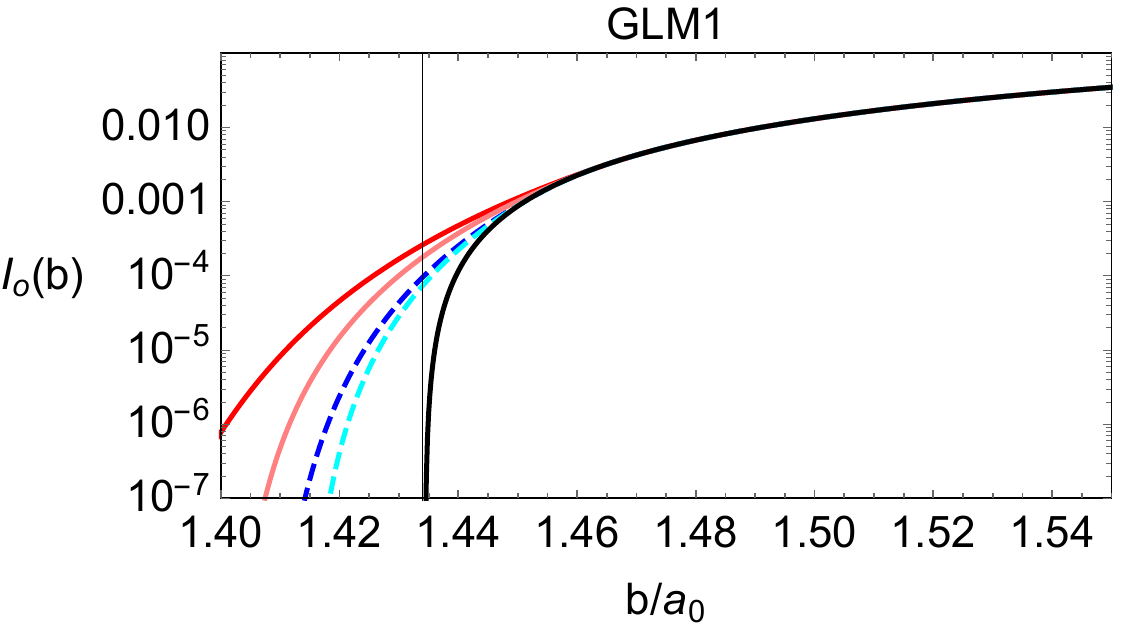}
 \includegraphics[scale=0.43]{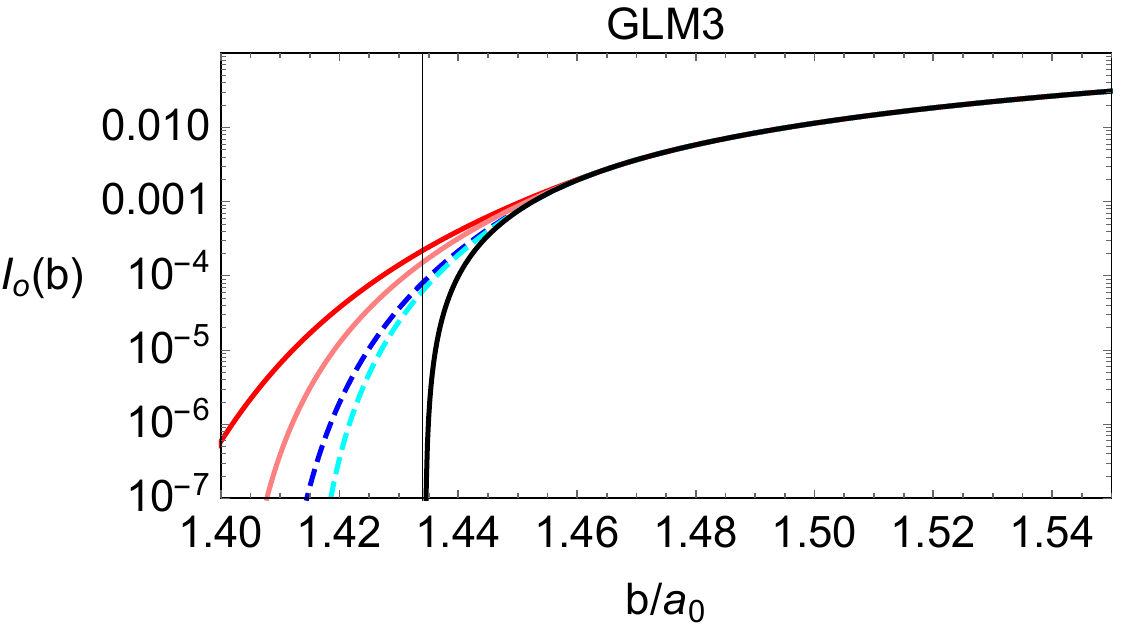}
\caption{The observed intensities near the inner shadow of the classical Schwarzschild black hole (black) and darkened QHCO images with $(k,\eta)=(1/20,1)$ (red), $(1/20,2)$ (pink), $(1/30,1)$ (dashed-blue), and $(1/30,2)$ (dashed-cyan). The upper and the lower panels correspond to the thin disk with GLM1 and GLM3 emission models, respectively. The QHCO images are slightly brighter than the classical black-hole ones. For a given $k$, increasing $\eta$ reduces the excess intensity, while increasing $k$ with $\eta$ fixed enhances it.}
\label{fig:darkintensity} 
\end{figure}

We demonstrate this expectation by the results in Fig.~\ref{fig:darkintensity}. 
The black curves show the observed intensity near the inner shadow of the classical Schwarzschild black hole image. The intensity drops to zero at $b/a_0=1.434$ (thin vertical lines) which corresponds to the direct emission of the event horizon. On the other hand, the colored curves represent the intensities of the darkened QHCO images with different values of $(k,\eta)$, showing that the darkened QHCO image is slightly brighter than the Schwarzschild image near the inner shadow. Here, the intensities of the direct emission outside $R_\textrm{out}$ overlap that of the Schwarzschild images. Moving inward from $R_\textrm{out}$, the redshift factor $|g_{tt}|$ of QHCOs drops more slowly than that of classical black holes. Therefore, the direct emission just inside $R_\textrm{out}$ appears slightly brighter near the inner shadow in the darkened QHCO images.

In addition, one can see from Fig.~\ref{fig:darkintensity} that increasing $\eta$ with $k$ fixed reduces the excess intensity. This matches the expectation that stronger internal interactions can further darken the images. Also, one can find that increasing $k$ with $\eta$ fixed enhances the excess intensity, whose implication will be discussed in sec.~\ref{sec:conclusion}.

Lastly, we can expect that such a tiny enhancement of the observed intensity near the inner shadow shall be robust for darkened QHCO images as long as there is emission at $r\lesssim R_\textrm{out}$, since the enhancement is not so sensitive to the details of the emission model as shown by the two panels of Fig.~\ref{fig:darkintensity}.

In summary, the darkened QHCO images can be almost indistinguishable from the images of the classical Schwarzschild black hole, albeit some tiny excess intensities will appear near the inner shadow of the darkened QHCO image. This is our prediction characterizing this model, which may be observed in the future. Of course, if there is no emission near the QHCO surface and the event horizon (e.g. the GLM2 emission), the darkened images would be completely the same as the Schwarzschild one, as one can see from the lower panel of Fig.~\ref{fig:imagedark}.

\section{Conclusions}\label{sec:conclusion}

The observed images of the supermassive black holes recently released by the Event Horizon Telescope collaboration display the image feature with a bright ring encircling a dark region. This image feature is consistent with the image of a classical black hole covered by event horizons in GR. However, such an image feature could also appear for horizonless compact objects as long as they are sufficiently compact. In particular, when the object is illuminated by an optically thin accretion disk, a bright ring that consists of higher-order images on top of the direct emission of the disk naturally appears when a set of unstable spherical photon orbits exist around the object. On the other hand, the central brightness depression emerges when there exist some mechanisms that can effectively darken the images.

In the context of black hole imaging which explores the true nature of black holes, we have analyzed the images of the quantum horizonless compact object (QHCO) model when illuminated by a geometrically and optically thin accretion disk. The QHCO represents a collection of highly excited quanta that forms a dense configuration, which can be interpreted as the final state of the gravitational collapse of matter according to the 4D semi-classical Einstein equation. In particular, the parameters $k$ and $\eta$ are related to the fundamental quantum properties near the Planck scale, i.e., the number of degrees of freedom $n$ and the microscopic interaction.

The QHCO images contain the outermost bright ring, which is common to the classical black hole images, because of the existence of the photon sphere outside the surface $R_\textrm{out}$. On the other hand, since the QHCO has no event horizon, there exist light rays that would enter the surface, propagate through the interior region, and then escape to infinity if one waited a sufficiently long time. The ideal images of QHCOs generated by including such trajectories are given by Figs.~\ref{fig:image201} and \ref{fig:image203}, which contain extra inner rings that are absent in the classical black-hole images. However, the exponentially strong redshift inside QHCOs significantly delays the propagation of these photons. Therefore, in a physically reasonable timescale (see the condition \eqref{dark_condition}), the contributions from those photons are removed, which then effectively darkens the images. As a result, the physical QHCO images, given by Fig.~\ref{fig:imagedark}, become nearly indistinguishable from those of classical black holes. We also find that, if the light source extends slightly inside the outer surface $R_\textrm{out}$, the observed intensity near the inner shadow will be slightly brighter than those of classical black holes, as shown in Fig.~\ref{fig:darkintensity}. With the future improvement of dynamic range \cite{Ayzenberg:2023hfw}, looking for such excess intensity near inner shadows could be a potential, albeit challenging, way to probe near-horizon quantum physics. 

Thus, the QHCO model predicts the images consistent with current observations, together with the slightly different features from those of the classical cases, and thus could be a candidate for quantum black holes.

Let us now discuss the implications of the obtained results to quantum gravity. First, we would like to stress that, although we have focused mainly on the darkening effects caused by the redshifts, another crucial darkening mechanism, i.e., the interactions between infalling photons and the microscopic structure of the QHCO, has also been partially considered through the parameter $\eta$. Effectively, for a fixed value of $k$, increasing $\eta$ (from $\eta=1$ to $\eta=2$) can be interpreted as including more interaction channels among the microstates. We find in Fig.~\ref{fig:darkintensity} that a larger $\eta$ makes the images closer to a classical black hole, which can be understood as a darkening effect due to stronger interactions. As pointed out in Refs.~\cite{Kawai:2015uya,Yokokura:2023wxp}, the thermodynamic equilibrium of the QHCO in the heat bath is rooted in the interactions among internal quanta. A more detailed understanding of such internal interactions may require investigations of quantum gravitational effects, since the energy scale inside the QHCO is close to the Planck scale, $\sim m_p/\sqrt{n}$.

Also, increasing $k$ enhances the excess intensity in Fig.~\ref{fig:darkintensity}. Here, the relations \eqref{sigma} and \eqref{k_def} indicate that the order of the magnitude of $k$ is determined by the number of degrees of freedom $n$ in the theory, and the energy scale $m_p/\sqrt{n}$ coincides with the maximum energy scale for the semi-classical approximation to hold \cite{Dvali:2007wp}. Therefore, $n$ should be the number of degrees of freedom in a theory that barely allows the semi-classical approximation and connects to a quantum-gravitational description. Thus, these discussions imply that the highly accurate observations of the intensity around the inner shadow in the future may probe the nature of interactions and the number of degrees of freedom in quantum gravity. 

In addition to image features, the horizonless structure of QHCOs may generate gravitational wave echos following the typical ringdown stages \cite{Cardoso:2016rao,Cardoso:2016oxy,Oshita:2018fqu}. Also, because QHCO models can be consistently generalized to dynamical pictures when evaporation processes are taken into account \cite{Kawai:2013mda,Kawai:2017txu}, it will be interesting to consider how the images evolve in time during the evaporation, at least adiabatically, in dynamical frameworks. Besides, in order to make the models more astrophysically relevant, extending the QHCO models to spinning cases is certainly an essential direction of research. We plan to report these interesting topics elsewhere in the future.

\acknowledgments
C.Y.C. is grateful to Daniel R. Mayerson for the fruitful discussions during the early development of this work. C.Y.C. is supported by the Special Postdoctoral Researcher (SPDR) Program at RIKEN.
Y.Y. is partially supported by Japan Society of Promotion of Science 
(Grants No.21K13929) and by RIKEN iTHEMS Program.

\bibliographystyle{mybibstyle}
\bibliography{bib}

\end{document}